\documentclass{article}

\usepackage{arxiv}

\usepackage[utf8]{inputenc} 
\usepackage[T1]{fontenc}    
\usepackage{hyperref}       
\usepackage{url}            
\usepackage{booktabs}       
\usepackage{amsfonts}       
\usepackage{nicefrac}       
\usepackage{subcaption}
\usepackage{microtype}      
\usepackage{lipsum}
\usepackage{graphicx}
\usepackage{amsmath}
\usepackage{mathrsfs}
\usepackage[noend]{algpseudocode}
\usepackage{algorithm}
\usepackage{multirow} 
\usepackage{mathtools}
\usepackage{bm}
\graphicspath{ {./images/} }

\title{A machine learning based material homogenization technique for in-plane loaded masonry walls}

\author{
 Alejandro Cornejo \\
  Civil and Environmental Engineering\\
  Universitat Politècnica de Catalunya\\
  International Centre for Numerical Methods in Engineering \\
  Campus Norte UPC, 08034 Barcelona, Spain \\
  \texttt{alejandro.cornejo.velazquez@upc.edu} \\
   \And
 Philip Kalkbrenner \\
  Civil and Environmental Engineering\\
  Universitat Politècnica de Catalunya\\
  Campus Norte UPC, 08034 Barcelona, Spain \\
  \texttt{philip.kalkbrenner@upc.edu} \\
  \And
 Riccardo Rossi \\
  Civil and Environmental Engineering\\
  Universitat Politècnica de Catalunya\\
  International Centre for Numerical Methods in Engineering \\
  Campus Norte UPC, 08034 Barcelona, Spain \\
  \texttt{riccardo.rossi@upc.edu} \\
  \AND
 Luca Pelà \\
  Civil and Environmental Engineering\\
  Universitat Politècnica de Catalunya\\
  Campus Norte UPC, 08034 Barcelona, Spain \\
  \texttt{luca.pela@upc.edu} \\
}

\begin{document}
\maketitle
\begin{abstract}
In recent years, significant advancements have been made in computational methods for analyzing masonry structures. Within the Finite Element Method, two primary approaches have gained traction: Micro and Macro Scale modeling, and their subsequent integration via Multi-scale methods based on homogenization theory and the representative volume element concept. While Micro and Multi-scale approaches offer high fidelity, they often come with a substantial computational burden. On the other hand, calibrating homogenized material parameters in Macro-scale approaches presents challenges for practical engineering problems. \\
Machine learning techniques have emerged as powerful tools for training models using vast datasets from various domains. In this context, we propose leveraging Machine Learning methods to develop a novel homogenization strategy for the in-plane analysis of masonry structures. This strategy involves automatically calibrating a continuum nonlinear damage constitutive law and an appropriate yield criteria using relevant data derived from Micro-scale analysis. The optimization process not only enhances material parameters but also refines yield criteria and damage evolution laws to better align with existing data.\\
To achieve this, a virtual laboratory is created to conduct micro-model simulations that account for the individual behaviors of constituent materials. Subsequently, a data isotropization process is employed to reconcile the results with typical isotropic constitutive models. Next, an optimization algorithm that minimizes the difference of internal dissipated work between the micro and macro scales is executed.\\
We apply this  technique to the in-plane homogenization of a Flemish bond masonry wall. Evaluation examples, including simulations of shear and compression tests, demonstrate the method's accuracy compared to  micro modeling of the entire structure.
\end{abstract}


\section{Introduction} \label{sec:intro}

Masonry is an assembly of many different materials - a conjunction of rocks/stones
or bricks usually bound together with mortar. There is no unique deﬁnition of how to allocate the units in order to obtain masonry, as it depends on the explicit knowledge developed by each cultural group in order to construct. Nowadays there is a large collection of  different masonry construction techniques. In any region, masonry features the local and traditional way of building. 
Many of the still existing structures around the globe belong to instances that represent  social values of humanity, \textit{e.g.}  churches, temples and mosques while palaces represent the administrative institutions that have allowed the development of our society. The great importance of the aforementioned constructions as well as the interest in maintaining structures of great cultural value entails the need for tools to model their behaviour and thus predict their health and future performance. 

Nowadays, demanding mathematical tasks are done by computers. In parallel with a continuous improvement in computer performance, numerical methods applied to the analysis of general structures have become more relevant due to their generality, accuracy and applicability in the field. The Finite Element Method (FEM) \cite{EO, OZ2} is one of the most renowned approaches employed to perform structural analysis. The FEM can then be utilized in order to compute the stress/deformations corresponding to particular loading and boundary conditions. A great visualization of the results is given by post processing software. The availability of such numerical tools has also brought the structural analysis of masonry structures to a next level, despite its analysis is complex and not yet entirely solved.  {Indeed, the analysis of non-linear orthotropic materials in which the tension and compression strengths are different, like concrete, it is still challenging. In the best case scenario, a large amount of material parameters need to be calibrated, limiting its application use.}

The availability of numerical models suited for masonry structures analysis is a fundamental step towards overcoming the limitations faced by the master builders of ancient times. In the past decades, several numerical modeling techniques for the assessment of existing masonry buildings have been developed. Masonry structures are more vulnerable when exposed to horizontal loading. Such loads can be induced by strong winds or earthquake actions. Under such conditions, masonry’s poor strength in tension can be exhausted easily and large cracks can lead to the total collapse of this kind of buildings.

The more accurate way of modeling masonry is the renowned micro modeling approach. Masonry is modeled by differing between the material components. Broadly spoken, the geometrical model consists of brick units and mortar joints. Distinct constitutive models assigned to each component are able to represent its respective natural material behavior \cite{page78}. The numerical simulation of such micro modeled structures is able to represent the damage and cracks very accurately when compared to experimental data. However, this approach requires a large computational cost and restricts its application to small-scale structures  {, especially in the case of the so called $\text{FE}^2$}. Besides the FEM, there are other numerical methods applied for the study of masonry like the Discrete Element Method (DEM) \cite{cundall, dem2, dem3, dem5, demThornton}, or even a combination between FEM and DEM in coupled methodologies \cite{femdem2d, femdem3d, femdemMMG, Cornejo_thesis, blast_jose_manuel} in which smeared continuum damage is employed and, after an element erosion method, a set of particles are generated through the calculated discrete crack. For a more complete description of the different procedures available the reader is referred to Lourenço \cite{Lourenço}, Roca \textit{et al.} \cite{roca2010} and D'Altri  \textit{et al.} \cite{altri}.  {The DEM is especially useful when modelling particle-assembled domains but in its bonded version can also be used to simulate bulk materials. The DEM reproduces naturally cracks through the domain by damaging and eventually removing the bonds between particles. One of the main intrinsic drawbacks of the DEM is that employs an explicit time advancing scheme, which implies the use of very low time steps, increasing dramatically the computational cost, especially for quasi-static simulations. Another important inconvenient is the large amount of particles needed to model a solid component, which also limits its applicability to real large scale cases.}

\begin{figure}
    \centering
    \includegraphics[width=1.0\textwidth]{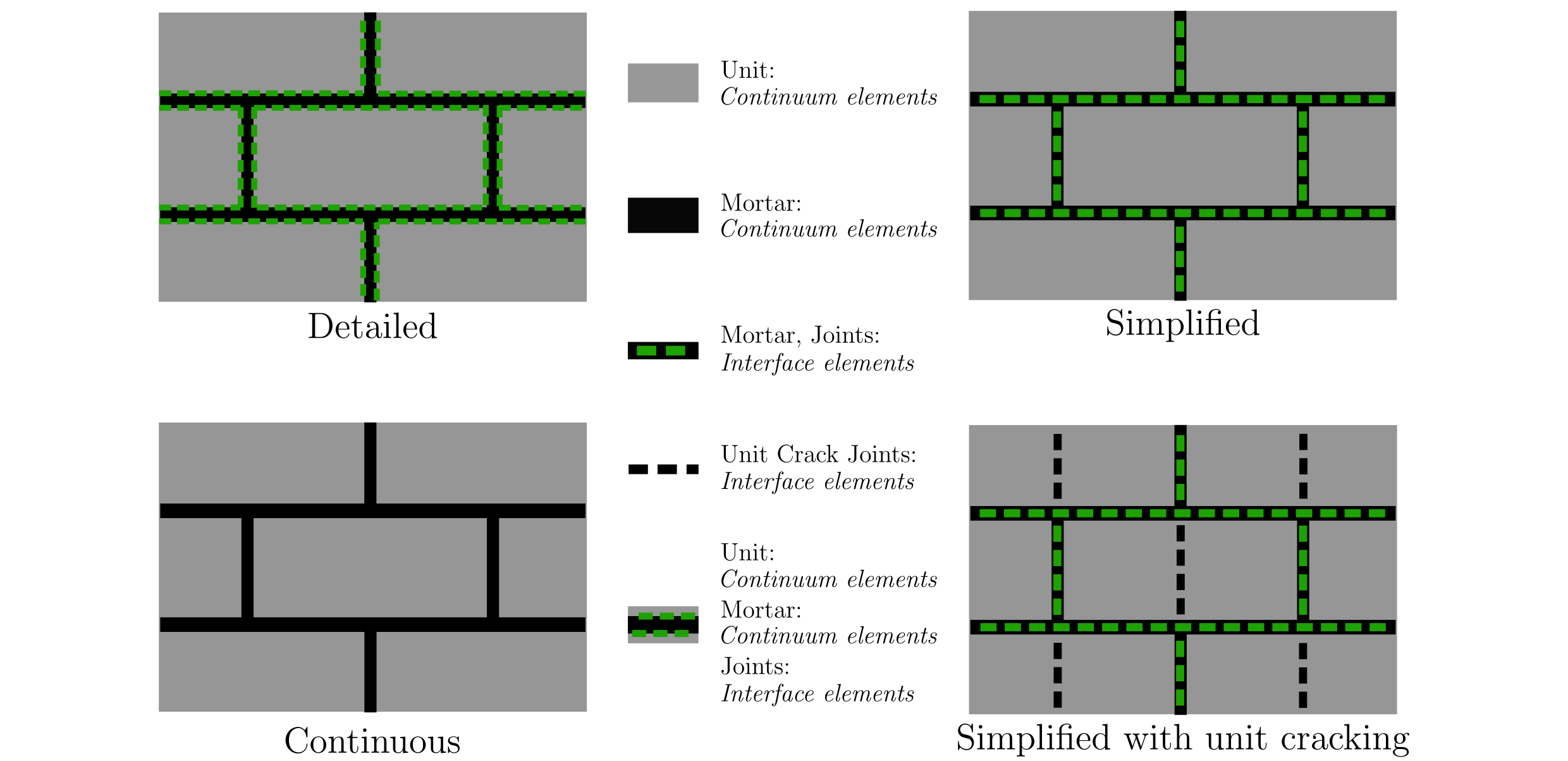}
    \caption{Overview of available micro modeling techniques for masonry. Showing the modeling background for the brick units, the mortar and a unit mortar joint interaction. Source: Kalkbrenner \cite{PhilippThesis}.}
    \label{fig:micro_models}
\end{figure}

The main aim of masonry micro models is the ability to model all possible failure modes of masonry structures. These modes have been presented carefully in numerous investigations like in Lourenço and Rots \cite{Lourenço93,Lourenço97} and in Petracca \cite{Petracca_thesis, Petracca2017} in which they try to model different crack mechanisms in masonry structures. Within the Micro approaches one can classify them depending on the level of detail achieved into the Detailed Micro Modelling (DMM), Simplified Micro Modelling (SMM) and Continuous Micro Modelling (CMM) as depicted in Fig. \ref{fig:micro_models}.

The DMM of masonry is the most accurate one. This technique takes into consideration the components and an interaction of the bond between the components while setting up the micro model. The brick units and the mortar joints are then discretized with continuum elements. Additional interface elements represent the frictional behavior in between the two components. The detailed approach allows an accurate response of the FE analysis results.  Nonetheless,  this procedure is highly expensive on computational cost. \\
The SMM ranges back to the research made by Page \cite{Page79}. Page assumed that the brickwork masonry consists of linear elastic brick units and inelastic mortar joint interactions. Back in the days, this procedure was not called SMM technique, since it was the ﬁrst general approach to model masonry by distinction of the units and the joints. Even though this technique is still in use, it is restricted to mortar joint cracking only and is not capable of representing all the cracking mechanisms of masonry. The reader is redirected to Lotfi and Shing \cite{lotfi}, Lourenço \cite{Lourenço97}, Lourenço and Rots \cite{Lourenço97}, Gambarotta and Lagomarsino \cite{Gambarotta97}, Macorini and Izzuddin \cite{Macorini11}, Petracca et al. \cite{Petracca17, Petracca_thesis} and Kumar and Barbato \cite{Kumar19} for more examples of application of the SMM techniques. \\
Finally, the CMM is considering the components of the masonry as continuum elements being connected without any interface interaction. Both the components are then assigned with a respective linear-elastic or nonlinear damage material model. For example, in Page \cite{Page88}, by performing experimental tests, the material behavior of the respective component has been investigated very properly and resulted in the deﬁnition of continuum models for each component. Further investigations and applications of the CMM techniques have been carried out in  Berto \textit{et al.} \cite{Berto04}, Barbosa \textit{et al.} \cite{Barbosa10}, Parisi \textit{et al.} \cite{Parisi11}, Drougkas \textit{et al.} \cite{Drougkas19}, Petracca \textit{et al.} \cite{Petracca_thesis, Petracca17},  among many more. Even obtaining good results, the above methodologies have a very high computational cost, which makes them impractical in large-scale cases. \\

Large investigative effort was put into the developments of techniques that assume masonry to be a continuum material with smeared material properties. This technique is called Macro modeling and it models the complex behaviour of masonry by means of a homogenized constitutive model that includes the material degradation \cite{VLACHAKIS2019534,Cervera10, Saloustros17, SaloustrosReview}, non-recoverable plastic deformations \cite{Saloustros17b} or even its anisotropic generalization \cite{Pel2011ContinuumDM,PELA2013957, PELA2014196}. The major advantages are less modeling and computational effort. Thus, large-scale structures can be analyzed through simulations of earthquakes, wind loads, soil settlements, etc. Many different types of simulations have been performed and are still being developed in order to investigate the causes of a wide variety of  structural problems. However, masonry is a highly anisotropic composite material and averaging its properties makes each analysis strongly dependent on the assumptions made by the analyst while deﬁning the homogenized constitutive model and its homogenized properties. Many research insights can be found in Lourenço \textit{et al.} \cite{Lourenço11}, Roca \textit{et al.} \cite{Roca13}, Saloustros \textit{et al.} \cite{Savvas15, Saloustros18a} and in Milani and Clementi \cite{Milani21}.
In order to overcome this in-accuracy, multi-scale approaches have been investigated as they are able to combine the micro and macro modeling. Such techniques are also known as $FE^2$ methods, since the boundary problem of the FE problem is solved at two scales with strongly differing sizes. A general introduction to the $FE^2$ methods is given in Schröder and Hackl \cite{Schroder14} and some applications in masonry in Massart \textit{et al.} \cite{Massart1,Massart2,Massart3}. These approaches have shown to be able to connect the advantages of both techniques, even though the computational cost is still too high. \\ 

It becomes clear from the previous considerations that innovative tools must be developed for the numerical analysis of masonry structures in order to minimize their current computational cost and optimize the assumed material properties at the same time at the macro-scale.  Machine learning (ML) has revolutionized scientific research by offering several significant advantages. One key advantage is its ability to handle vast amounts of complex data with remarkable efficiency. By utilizing powerful algorithms, machine learning enables researchers to analyze and extract patterns from massive data-sets, providing valuable insights and facilitating data-driven decision-making. Additionally, machine learning models possesses the capability to learn and improve over time, continuously refining their performance based on experience. This iterative learning process empowers scientists to tackle intricate problems and optimize their experimental approaches. Furthermore, machine learning techniques have the potential to discover novel correlations and relationships within data, opening up new avenues for scientific exploration. 

The use of machine learning for the structural design has been driven forward by Adeli and Yeh \cite{Adeli89}). In this research a perceptron based machine learning model has been trained to design structural steel beams. At that time training set sizes were restricted due to computation power. Very good insights of machine learning in structural engineering are given in Amezquita-Sanchez \textit{et al.} \cite{Amezquita20}. Many of the current investigations are focused on structural system identiﬁcation by machine learning and Artificial Neural Networks (ANN) \cite{Jiang16, Youngding18}. Other researchers investigate on structural health monitoring and use machine learning to analyze crack propagation \cite{Ibrahim19} or damage detection in bridges submitted to temperature gradients \cite{Kostic17}. \\

As mentioned before, the research ﬁeld of masonry and heritage structures has gained importance in the last decades. Therefore, novel numerical tools such as ANNs have been also applied to improve the assessment methods for masonry buildings. Garzón-Roca \textit{et al.} \cite{Garzon13} investigated the use of an ANN to determine the maximum axial load that can be withstood by a masonry wall, and Plevris and Asteris \cite{Plevris14} developed a method to predict the yield criteria for the anisotropic failure surface of masonry. \\

In light of the aforementioned factors, the field of numerical analysis for masonry structures faces two prominent challenges. First, there is a need to develop advanced numerical simulation techniques capable of accurately evaluating the diverse structural response exhibited by large masonry structures. These techniques must be able to effectively capture the intricate behaviors and complexities inherent in masonry systems. Second, constitutive models utilized in macro modeling approaches must be carefully defined to ensure an accurate representation of the unique material behavior exhibited by masonry. These models should account for the specific characteristics and mechanical properties of masonry materials, allowing for reliable predictions of structural performance. Addressing these challenges is crucial for enhancing the accuracy and reliability of numerical analyses in the field of masonry structures. \\

Accordingly, this paper endeavors to tackle both of these challenges comprehensively. Firstly, it focuses on evaluating the potential of macro modeling by introducing a novel nonlinear constitutive law specifically designed for the seismic analysis of irregular masonry structures. The technique is applied to a complex case study to assess its efficiency and accuracy. Through this application, the conventional approach of defining macro model properties is examined, revealing its limitations. Secondly, the research investigates the integration of ML to address and overcome some of these limitations. The primary objective is to leverage the capabilities of ML as a modern tool to develop well-trained constitutive models using data obtained from micro modeling at a smaller scale. This approach leads to the development of a novel homogenization strategy, utilizing ML to reliably calibrate relevant parameters for macro modeling of masonry structures. By incorporating ML, this research aims to enhance the accuracy and effectiveness of modeling masonry structures, thereby contributing to advancements in the field. \\

This paper is organized as follows: in Section \ref{sec:CL}, the theoretical foundations of the constitutive models employed for the nonlinear analysis of masonry are thoroughly explained. The selected continuum damage law relies on two distinct damage variables, each representing the extent of damage in tension or compression. These damage variables follow separate evolution laws, capturing the progression of damage under different loading conditions. In Section \ref{sec:ML}, a novel homogenization technique based on ML is presented. The Section begins by providing essential background information, including the utilization of numerical software tools, gradient descent operators, loss functions, and the computation graph, to establish the foundation for the proposed approach. Additionally, the concept of a virtual laboratory is introduced, emphasizing the generation of extensive data sets comprising coupled strain and stress pairs through numerous micro scale analyses on masonry representative volume elements. These data sets are subsequently employed as training input for the ML model. By leveraging the power of ML, the proposed technique offers a promising avenue for accurately predicting the behavior of masonry structures by effectively capturing their complex material response. Section \ref{sec:ML} thus contributes to the field by combining the principles of ML and numerical analysis, leading to the development of a robust and efficient homogenization strategy for masonry structures. To showcase the functionality and versatility of the approach, several numerical benchmark examples are presented in Section \ref{sec:example}, demonstrating its effectiveness in accurately simulating masonry type of structures. Furthermore, the constitutive model is applied to a masonry wall, and the resulting numerical analysis results are compared with experimental tests, affirming the accuracy and reliability of the model. Finally, in Section \ref{sec:conclusions} some conclusions are drawn in view of the results obtained and some future works are stated.

\section{Constitutive Model} \label{sec:CL}

\subsection{Constitutive law}\label{const_law}

In this paper, the developed constitutive law is based on \textit{damage} type degradation material models. In these models, we assume that there is a relationship between the damaged and undamaged (effective) stress state such as:

\begin{equation}
    \Bar{\boldsymbol{\sigma}} = \textbf{C} : \boldsymbol{\varepsilon}
    \label{eq:effective_stress}
\end{equation}

\noindent
where $ \Bar{\boldsymbol{\sigma}}$ is the effective stress tensor, \textbf{C} the elastic constitutive tensor and $\boldsymbol{\varepsilon}$ the total strain tensor. The works of Cervera \textit{et al.} \cite{CerveraDplus} , Faria \textit{et al.} \cite{FariaDplus} and Wu \textit{et al.} \cite{WuDplus} introduced a split of the internal damage variables into de tensile ($d^+$) and its compressive ($d^-$) counterparts. Based on this split, the integrated stress tensor could be calculated as

\begin{equation}
    \boldsymbol{\sigma} = (1-d^+)\, \Bar{\boldsymbol{\sigma}}^+ + (1-d^-) \,\Bar{\boldsymbol{\sigma}}^-
    \label{eq:stress_tensor}
\end{equation}

\noindent
where, as mentioned before, $d^+$ and $d^-$ represent the degradation achieved due to tensile and compressive loads, respectively; each of the damages corresponds to a scalar value ranging from 0 (intact) to 1 (fully damaged). Additionally, the $\Bar{\boldsymbol{\sigma}}^+$ and $\Bar{\boldsymbol{\sigma}}^-$ represent the spectral decomposition of the total stress in Eq. \eqref{eq:effective_stress} performed according to Wu \textit{et al.} \cite{WuDplus} as

\begin{equation}
    \Bar{\boldsymbol{\sigma}}^+ = \textbf{P}^+ : \Bar{\boldsymbol{\sigma}}
    \label{equ:effective_stress_composition_pos}
\end{equation}
\noindent
and

\begin{equation}
    \Bar{\boldsymbol{\sigma}}^- = \Bar{\boldsymbol{\sigma}}-  \Bar{\boldsymbol{\sigma}}^+= \textbf{P}^- : \Bar{\boldsymbol{\sigma}}
    \label{equ:effective_stress_composition_neg}
\end{equation}

\noindent
where $\textbf{P}^+$ and $\textbf{P}^-$ are fourth order projection tensors expressed as \cite{FariaDplus}

\begin{equation}
    \textbf{P}^+ = \sum_{i=1} H(\Bar{\sigma}_i) \textbf{p}_{ii} \otimes  \textbf{p}_{ii}
\end{equation}

\begin{equation}
    \textbf{P}^- = \textbf{I} - \textbf{P}^+
\end{equation}

\noindent
where $\Bar{\sigma}_i$ is the \textit{i}-th principal stress (eigenvalue) of the effective stress tensor $\Bar{\boldsymbol{\sigma}}$. The symmetric tensor $\textbf{p}_{ii}$ is the outer product of the eigenvector $\textbf{n}_i$ corresponding to the \textit{i}-th principal stress. The Heaviside function $H(\Bar{\sigma}_i)$ guarantees that the positive projection tensor $\textbf{P}^+$ is computed by using only the positive values of the effective stress tensor; it is defined as

\begin{equation}
H(x) = 
\left\{
\begin{array}{ccc}
        0, \,\, \text{if}\,\, x \leq 0 \\
        1, \,\, \text{if}\,\, x > 0
\end{array}
\right.
\end{equation}

\subsection{Yield criteria} \label{yield} 

In order to identify the initiation of the material degradation in both tensile and compressive states, a proper yield criterion must be stated in each case. In Lubliner \textit{et al.} \cite{plasticdam} a well suited yield criteria is introduced that considers two scalar values that follow different material properties in tension and compression. The values $\tau^+$ (tension) and $\tau^-$ (compression) represent the scalar uni-axial equivalent stress of the respective effective stress counterparts $\Bar{\boldsymbol{\sigma}}^+$ or $\Bar{\boldsymbol{\sigma}}^-$, respectively. The uni-axial equivalent stresses are computed according to the modifications introduced in Petracca \textit{et al.} \cite{Petracca_thesis} as

\begin{equation} \label{tau_plus}
    \tau^+ = H(\Bar{\sigma}_{max}) \left[   \frac{1}{1-\alpha} (\alpha\, \Bar{I}_1+\sqrt{3\Bar{J}_2} +\beta\, \Bar{\sigma}_{max}) \frac{f_p^+}{f_p^-}\right]
\end{equation}

\begin{equation} \label{tau_minus}
    \tau^- = H(-\Bar{\sigma}_{min}) \left[   \frac{1}{1-\alpha} (\alpha\, \Bar{I}_1+\sqrt{3\Bar{J}_2} +\kappa \, \beta\, \left< \Bar{\sigma}_{max} \right>) \right]
\end{equation}

In Eqs. \eqref{tau_plus}-\eqref{tau_minus}, $\Bar{I}_1$ is the first invariant of the effective stress tensor (tensile or compressive counterpart), $\Bar{J}_2$ the second invariant of the deviatoric effective stress tensor and $\Bar{\sigma}_{max}$, $\Bar{\sigma}_{min}$ the maximum and minimum principal stresses of the effective stress tensor, respectively. The $\left< \cdot \right>$ represent the widely known Macaulay brackets. The constants $\alpha$ and $\beta$ are dimensionless scalar values and can be computed by the following expressions considering the yield values of tension and compression behavior of the material:

\begin{equation}
    \alpha = \frac{k_b-1}{2k_b-1}
\end{equation}

\begin{equation}
    \beta = (1-\alpha) \frac{f_p^-}{f_p^+}- (1+\alpha)
\end{equation}

\noindent
where $k_b$ is a multiplier to respect the increasing strength under bi-axial compression states, so that $k_b = \frac{f_{bi}^-}{f_p^-}$ with $f_{bi}^-$ being the bi-axial compression strength and $f_p^-$ the uniaxial peak strength in compression. $f_p^+$ corresponds to the tension peak strength of the material.

\begin{figure}
    \centering
    \includegraphics[width=0.4\linewidth]{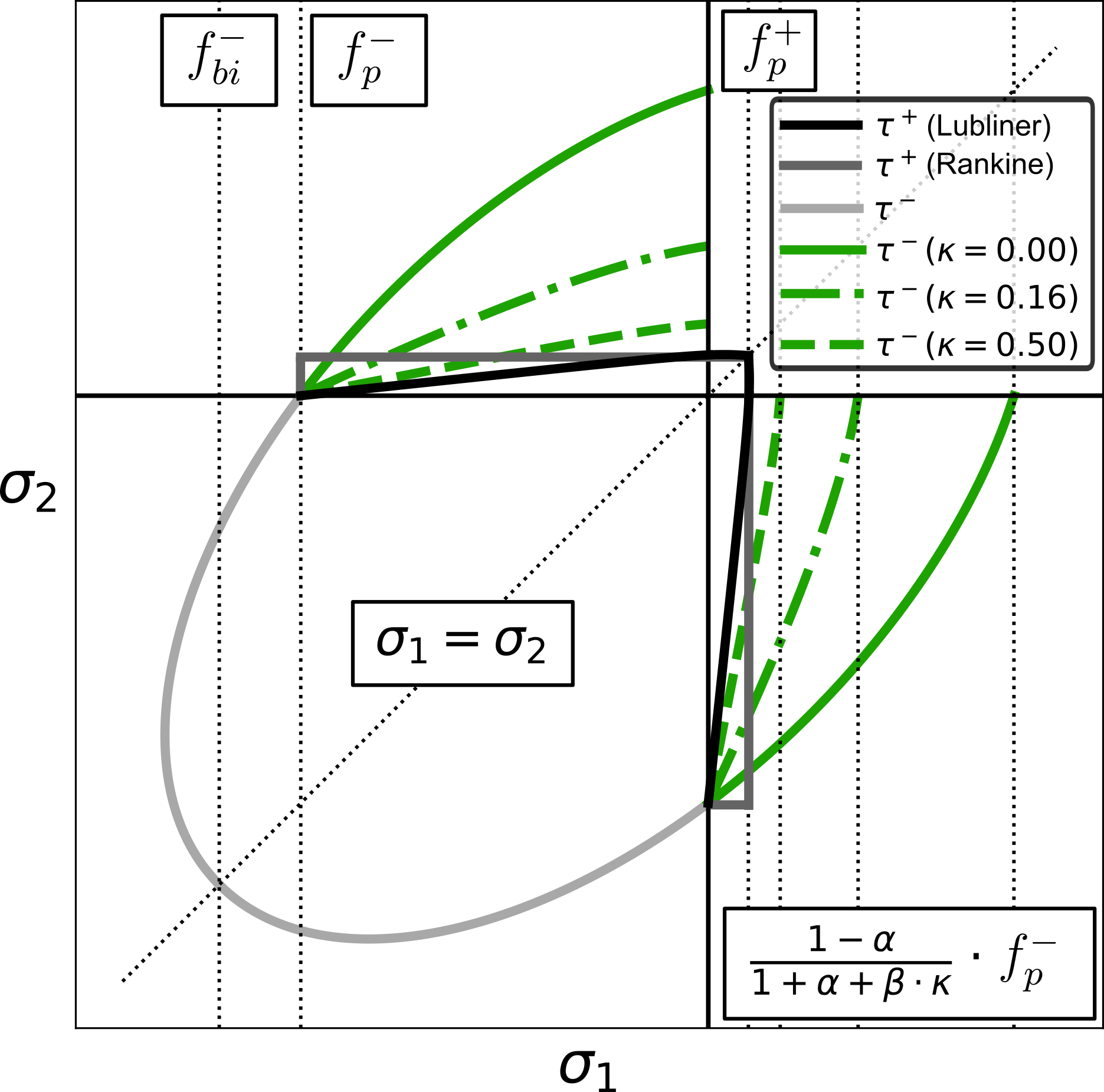}
    \caption{Yield surfaces for the tension equivalent stress $\tau^+$ and the compression equivalent stress $\tau^-$ in the two dimensional principal stress state. Original source Kalkbrenner \cite{PhilippThesis}.}
    \label{fig:plot_yields}
\end{figure}

Figure \ref{fig:plot_yields} shows the yield surface of the Rankine and Lubliner model in tension and the modiﬁed Lubliner model in compression in the two dimensional principal stress state, respectively. The effect of a variation of $k$, introduced by Petracca \textit{et al.} \cite{Petracca_thesis} in order to modify the surface in case of mixed stress states, is shown for three different values. The depicted yield surfaces are utilized in the application examples of this paper.

The expressions \eqref{tau_plus}-\eqref{tau_minus} are valid for 2D analysis. In case 3D analysis are considered, the negative equivalent stress must be extended inside the brackets by a coefficient that accounts for three dimensional compression stress effects \cite{Petracca_thesis}.

\subsection{Damage evolution} \label{sec:damage}

The evolution of damage is intrinsically related to the stress state and the strength of the material. Damage is an irreversible process and this is reflected in the model by storing the material threshold as a historical variable through all the process. Thus, two additional scalar values $r_n^{\pm}$ are introduced as actual threshold values and deﬁned as follows:

\begin{equation} \label{scalar_damages}
    r^{\pm} = max \left\{ r_0^{\pm}; \underset{t_0\leq n\leq e}{ \text{max}} \tau_n^{\pm} \right\}
\end{equation}

\noindent
where $n$ indicates the current step of an analysis going from a start time $t_0$ to an end time $t_e$. Furthermore, $\tau_n^\pm$ is the current uniaxial equivalent stress, $r_0^\pm$ is the initial threshold value, with $r_0^+=f_p^+$, if no hardening function in tension is considered and $r_0^+ = f_0^+$ , if a damage onset stress for tension behavior is considered. $r_0^- = f_0^-$ deﬁnes the elastic limit in compression. Finally, the yield function can be written as \cite{Petracca_thesis}:

\begin{equation}
    \Phi^\pm (\tau^\pm, r^\pm) = \tau^\pm- r^\pm \leq 0.
\end{equation}

Since the behaviour of masonry when subjected to tensile and compressive states is notably different, the damage evolution laws describing theses processes have been implemented according to different expressions as detailed in Petracca \textit{et al.} \cite{Petracca17}.

\subsubsection{Compression behavior} \label{subsubsec:compression_behavior}

To ensure hardening and softening in compression behavior, Petracca \textit{et al.} \cite{Petracca17} established a novel hardening-softening law based on quadratic B\'ezier curves. Figure \ref{fig:compressive_S_E} shows the thoroughly developed post linear evolution behavior. It consists of three Bézier splines and a final residual part. The computation of each region can be done as:

\begin{figure}
    \centering
    \begin{subfigure}{0.4\textwidth}
        \centering
        \includegraphics[width=\textwidth]{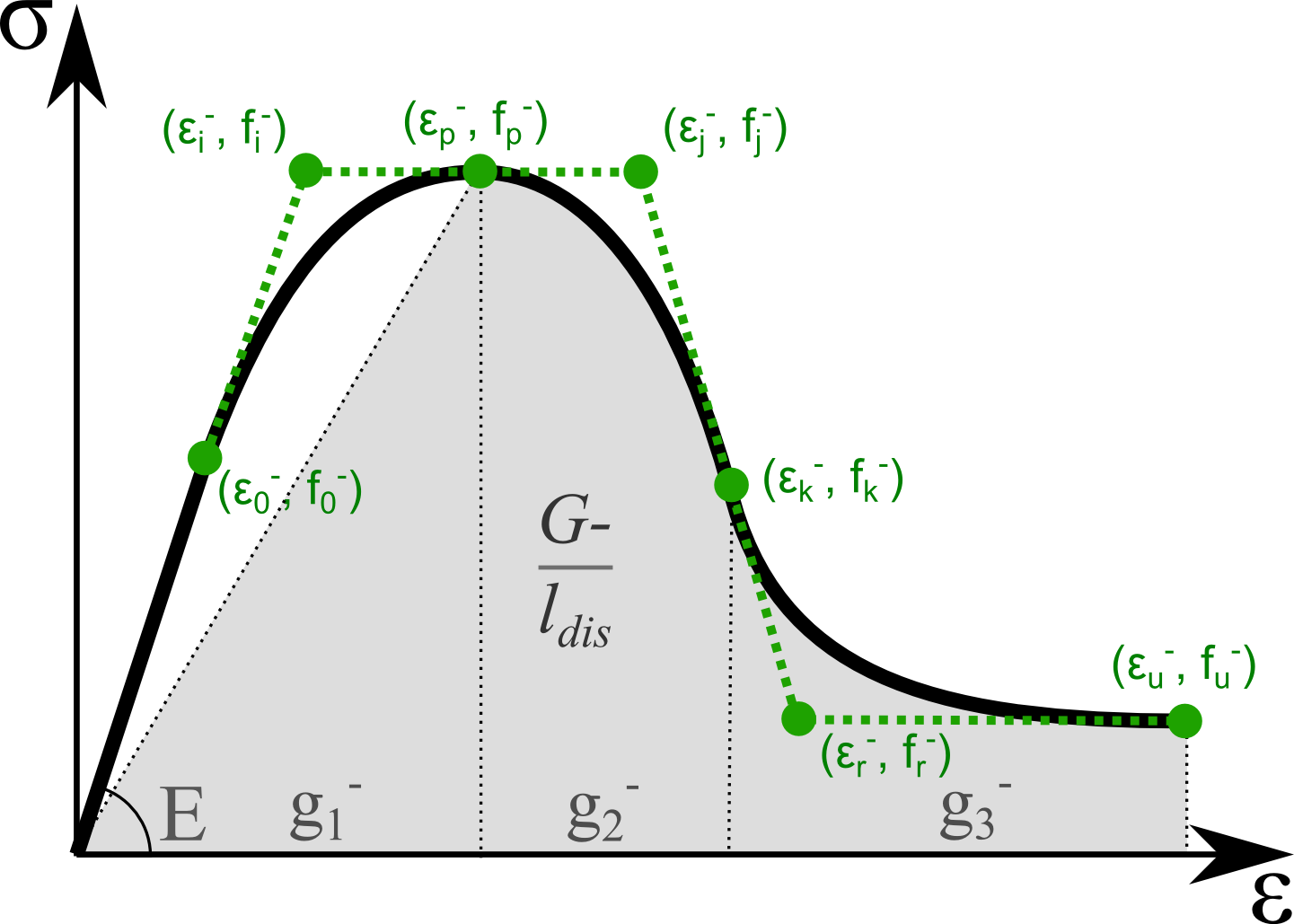}
        \caption{Uni-axial strain-stress curve of the material’s compression behavior with B\'ezier control nodes \cite{PhilippThesis}.}
        \label{fig:compressive_S_E}
    \end{subfigure}
    \hfill
    \begin{subfigure}{0.4\textwidth}
        \centering
        \includegraphics[width=\textwidth]{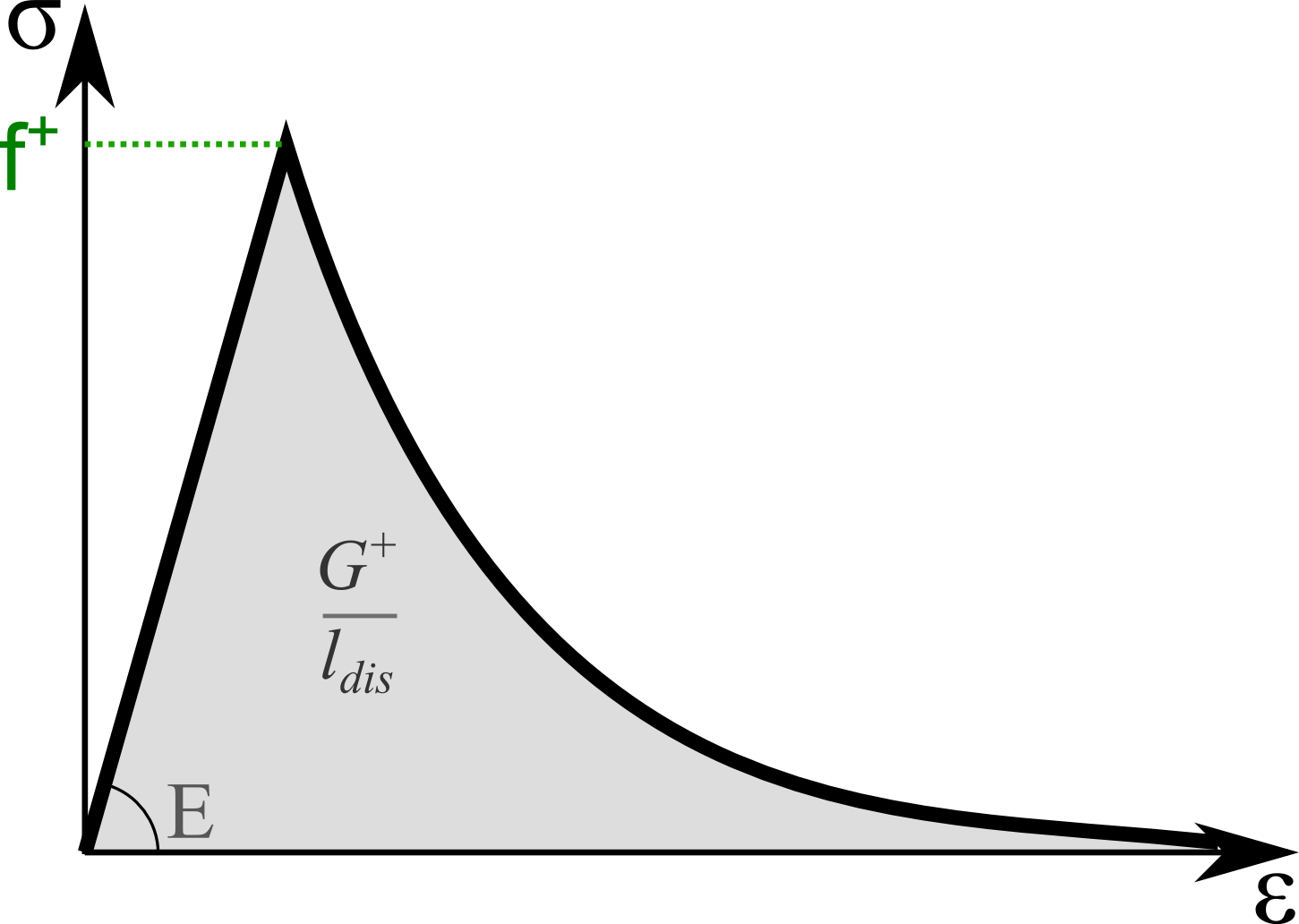}
        \caption{Uni-axial strain-stress curve of the material’s tension behavior \cite{PhilippThesis}.}
        \label{fig:tension_S_E}
    \end{subfigure}
    \caption{Different available hardening/softening stress-strain curves.}
    \label{fig:mainfigure}
\end{figure}

\begin{equation} \label{equ:bezier_splitting}
   \Psi(\xi^-) = 
\left\{
\begin{array}{ccc}
        \mathcal{B}(\xi^- , \varepsilon_0^- , \varepsilon_i^- , \varepsilon_p^- , f_0^- , f_i^- , f_p^- ) & \varepsilon_{c,0} < \xi^- \leq \varepsilon_p^- \\ \\
        \mathcal{B}(\xi^-, \varepsilon_p^-, \varepsilon_j^-, \varepsilon_k^-, f_p^-, f_j^-, f_k^-) & \varepsilon_p^- < \xi^- \leq \varepsilon_k^- \\ \\
        \mathcal{B}(\xi^- , \varepsilon_k^- , \varepsilon_r^- , \varepsilon_u^- , f_k^- , f_r^- , f_u^- ) & \varepsilon_{k} < \xi^- \leq \varepsilon_u^- \\ \\
        f_u^-
\end{array}
\right.
\end{equation}

\noindent
where $\mathcal{B(\cdot)}$ is a B\'ezier function that requires the coordinates of three control points and is deﬁned as

\begin{equation}
    \mathcal{B}(X, x_1, x_2, x_3,y_1,y_2,y_3) = (y_1-2y_2+y_3)p^2+2p(y_2-y_1)+y_1
\end{equation}

\noindent
with 

\begin{equation}
    p = \frac{-B + \sqrt{D}}{2A}
\end{equation}

\noindent
and

\begin{align}
    A &= x_1-2x_2+x_3 \\
    B &= 2(x_2-x_1) \\
    C &= x_1 -X \\
    D &=B^2-4A\,C.
\end{align}

$\xi^-$ is introduced as a strain-like counterpart to the current damage threshold $r^-$ and is deﬁned as the ratio of the yield strength and the materials Young’s ($E$) as $\xi^- = \frac{r^-}{E}$.

Finally, the compressive damage variable $d^-$ can be calculated according to:

\begin{equation} \label{damage_compression}
    d^- = 1-\frac{\Psi(\xi)}{r^-}
\end{equation}

A proper fracture energy regularization with respect to the finite element size must be performed in order to dissipate the expected amount of fracture energy ($g_1^-+g_2^-+g_3^-$ in Fig. \ref{fig:compressive_S_E}). A detailed discussion in this matter can be found in Petracca \textit{et al.} \cite{Petracca17, PhilippThesis}.  {This energy regularization stretches the softening part of the compression curve in order to satisfy a predefined energy density dissipation.}

\subsubsection{Tension behavior}

The tension behaviour is based on an exponential softening curve, hence, the damage variable $d^+$ is computed as:

\begin{equation} \label{damage_tension}
    d^+(r^+) = 1-\dfrac{r_0^+}{\tau^+}\mathrm{exp}\left(A\left(1-\dfrac{\tau^+}{r_0^+}\right)\right)
\end{equation}

\noindent
in which the $A$ parameter is determined from the energy dissipated in an uniaxial tension test as

    \begin{equation} 
    A= \left(  \dfrac{G_{f}E}{\hat{l}_{dis}f_{t}^{+,2}}-\dfrac{1}{2} \right)^{-1} \ge 0
    \label{char}
    \end{equation}

being $f_{t}^+$ is the tensile strength, $G_{f}$ is the specific fracture energy per unit area (taken as a material property) and $\hat{l}_{dis}$ is the characteristic length of the element. By using this characteristic length normalization, the size-effect problem is overcome.

For a consistent linearization of the constitutive model, a numerical estimation of the constitutive tangent tensor is performed via a perturbation method as described in Cornejo \cite{Cornejo_thesis}.

\section{Machine learning material homogenization} \label{sec:ML}

\subsection{Introduction} \label{sec:ML_introduction}
The idea of the present research is to avoid switching around at multiple scales (micro and macro) like it was proposed by Petracca \textit{et al. }\cite{Petracca_thesis, petracca2016a}. The key issue in this work is then finding a single homogeneous continuum damage model for the macro scale analysis of masonry that takes into account more realistically its anisotropic behavior due to the material heterogeneity of the micro scale. In this regard, and based on the prior work of Kalkbrenner \cite{PhilippThesis}, we propose to use a machine learning model to predict such a homogeneous continuum damage model
for masonry structures. In Kalkbrenner they defined a supervised training strategy based on the Adam optimizer.

\begin{figure}
    \centering
    \includegraphics[width=0.9\linewidth]{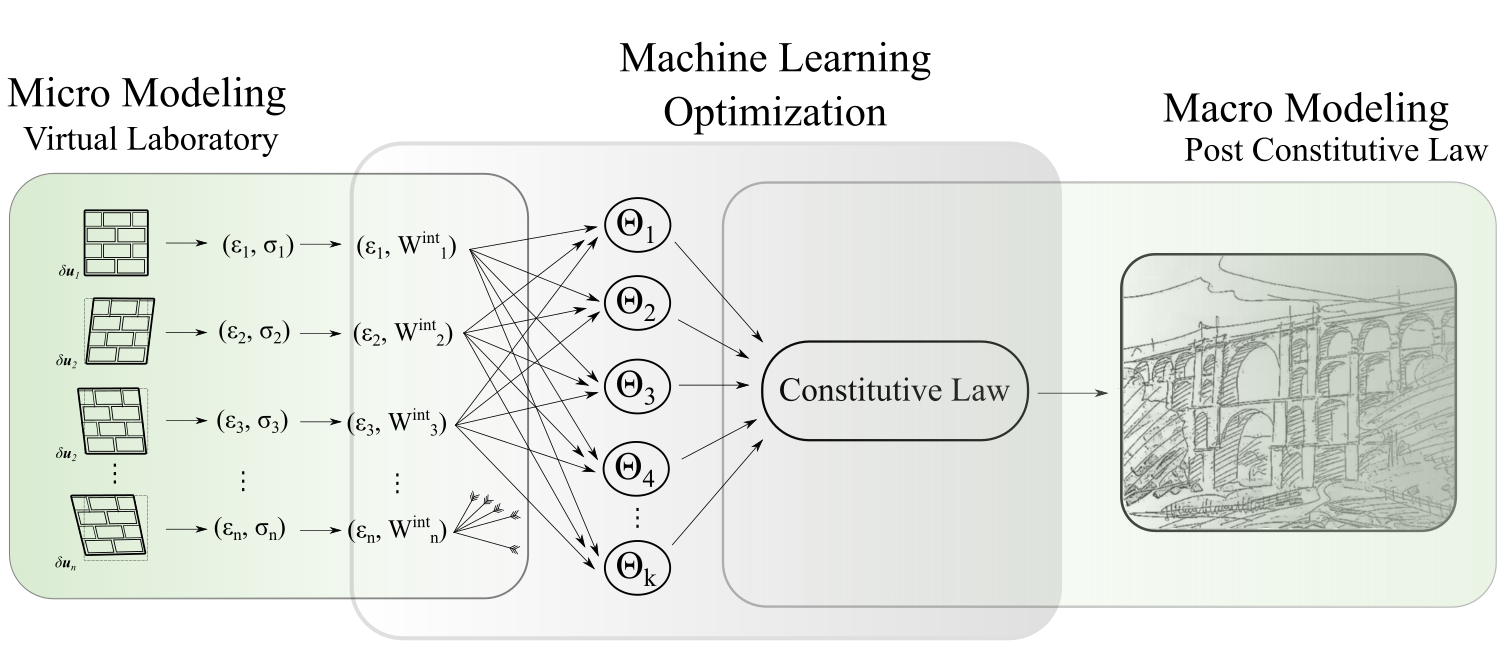}
    \caption{Overview of the machine learning homogenization technique, showing all included procedures at multiple scales and the machine learning technique as connector.}
    \label{fig:ML_overview}
\end{figure}

Fig. \ref{fig:ML_overview} shows the workflow of the proposed homogenization technique. It describes the main idea of producing data from a virtual laboratory at the micro scale, goes further to the machine learning and finalizes in a post machine learning constitutive law applicable to the analysis of masonry structures at the macro scale.

This section introduces all the steps required for the development of a machine learning model able to homogenize the complex masonry material at a macro scale level. Section \ref{sec:ml_cl_model} introduces the machine learning model applied in this research. It also serves as the connector of the analysis at the multiple scales. More specifically, this section details how the constitutive model from Section \ref{sec:CL} is integrated inside the machine learning framework. Next, Section \ref{sec:virtual_lab} introduces a space, where extensive and relevant data is obtained from the virtual laboratory; this information will be isotropized based on the work of Norris \cite{Norris06} and Rossi \textit{et al.} \cite{Rossi21}. The proposed isotropization procedure becomes necessary, since the data coming from the virtual laboratory are following an anisotropic behaviour, while the constitutive law integrated into the machine learning is based on the hypothesis of isotropic elasticity.

After having isotropized the data, the machine learning model can be trained based on several virtual laboratory calculations of an RVE of different representative strain fields.

\subsection{The constitutive law machine learning model} \label{sec:ml_cl_model}

In this paper, our objective is to establish a connection between strains and stresses for a homogenized material. The relationship we seek can be theoretically described by a constitutive law applicable to any material. Hence, we derive the mathematical formulation for our machine learning model based on this constitutive law, defining it as the function $\Psi(\boldsymbol{\varepsilon}_{true})$. The input and output for the machine learning process consist of a collection of strain histories $\boldsymbol{\varepsilon}_{true}$ and their corresponding stresses $\boldsymbol{\sigma}_{true}$, which can be transformed in a strain-dissipated work ($\boldsymbol{\varepsilon}_{true}, W_{int}$) pairs data set. To obtain the predicted output of the machine learning model, we adjust a set of specific constitutive law parameters denoted as $\boldsymbol{\Theta}$. Any strain-stress history can be transformed to a more meaningful and manageable strain-dissipated work by means of:

\begin{equation}
    W_{int} =  \int_0^{t_f} \int_\Omega \boldsymbol{\sigma}:\Dot{\boldsymbol{\varepsilon}} \, d\Omega\, dt.
    \label{work_calculation}
\end{equation}
\noindent where $t_f$ is the final time of the simulation and $\Omega$ the solid domain.

A predeﬁned learning criterion checks the error value of the actual optimization state according to a loss function $\mathcal{L}$. If the error value is less or equal to the learning criterion, the optimization ﬁnishes and stores the actual modiﬁcation of the parameters $\boldsymbol{\Theta}$ as the optimized model parameters $\Tilde{\boldsymbol{\Theta}}$. Then the model can be utilized to predict stresses $\Tilde{\boldsymbol{\sigma}}_{pred}$ by entering any strain state $\Tilde{\boldsymbol{\varepsilon}}$ as

\begin{equation}
    \Psi(\Tilde{\boldsymbol{\varepsilon}}, \Tilde{\boldsymbol{\Theta}}) = \Tilde{\boldsymbol{\sigma}}_{pred}.
\end{equation}

\noindent where its corresponding predicted internal work $\tilde{W}_{int}$ can be also computed according to Eq. \eqref{work_calculation}.

The cost function $\mathcal{L}$ to be optimized with respect to the parameters $\boldsymbol{\Theta}$ is built as

\begin{equation}
    \mathcal{L} (\boldsymbol{\Theta}) = \sum_{k=1}^{n}\left|(\tilde{W}_{int,k}(\boldsymbol{\Theta}) - W_{int,k})\right|
    \label{cost_L}
\end{equation}

\noindent
where $n$ is the number of strain-stress reference histories. Qualitatively, the cost function is the summation of the difference of internal work dissipated in all load histories.

\begin{figure}
    \centering
    \includegraphics[width=0.9\linewidth]{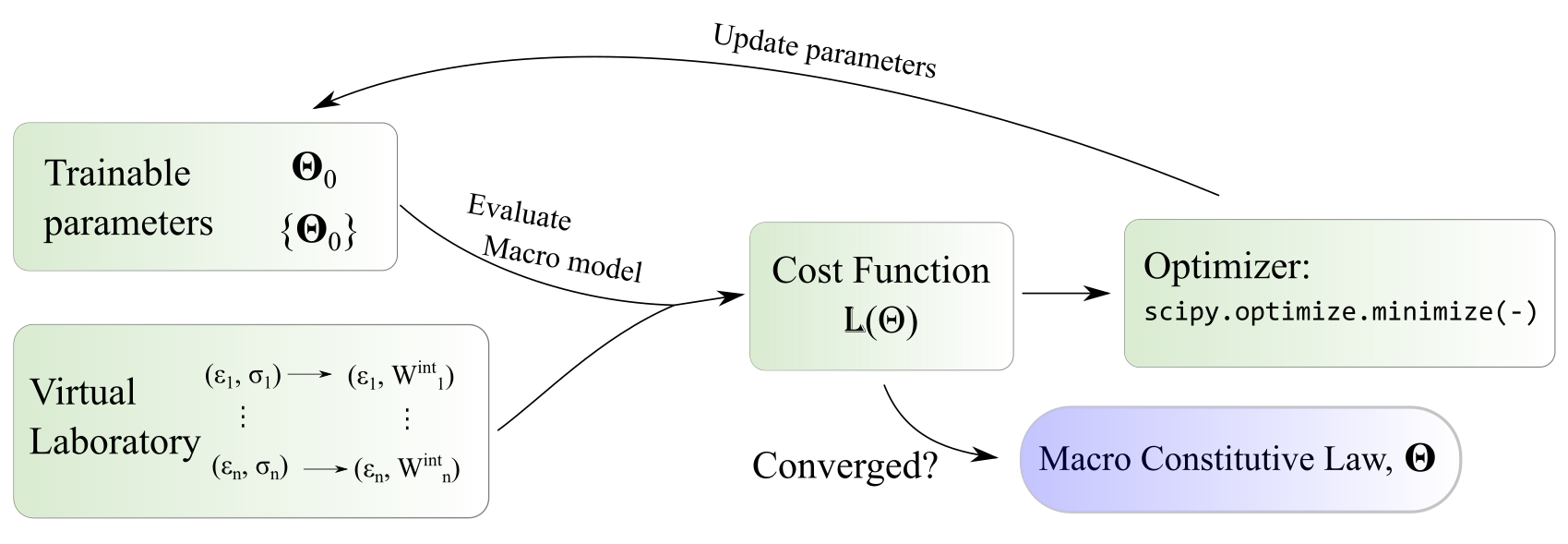}
    \caption{Parameter optimization procedure $\mathcal{P}$.}
    \label{fig:parameter_opt}
\end{figure}

This output is then passed to the optimization algorithm $\mathcal{P}$. The chosen optimizer technique is a trust region method \cite{trustmethods} implemented within the Python library Scipy \cite{2020SciPy-NMeth}. This methodology minimizes the values of the cost function $\mathcal{L} (\boldsymbol{\Theta})$ by  updating a set of parameters subjected to a user 
defined initial guess ($\boldsymbol{\Theta}_0$) and bounds (\{$ \boldsymbol{\Theta} $\}). This procedure is schematically depicted un Fig. \ref{fig:parameter_opt}.

Algorithm \ref{algorithm:machine_learning_optimization_procedure} presents the  implementation of the optimization procedure of the machine learning homogenization technique. It includes the building of the database, which is done once at the beggining of the calculation, and the iterative optimization of the parameters that minimize the proposed cost function.

\begin{algorithm}[]

\caption{Find $\bm{\Theta}^\star$ so that $\bm{\Psi}\Big(\bm{\varepsilon}_{true}, \bm{\Theta}^\star \Big) \approx \bm{\sigma}_{true}$}
\label{algorithm:machine_learning_optimization_procedure}
\hspace*{\algorithmicindent} \textbf{input:} $\bm{\varepsilon}_{true}$, $\bm{\sigma}_{true}$, $\bm{\Theta}_{0}$, bounds of $\bm{\Theta}$,  $\texttt{n}_{epoch}, tol$\\
\hspace*{\algorithmicindent} \textbf{output:} $\bm{\Theta}^\star$\\
\begin{algorithmic}[1]
		\State Build reference placeholders $\hat{\bm{\varepsilon}}$, $\hat{\bm{\sigma}}$, $\hat{W}_{int}$
		\State Build variable list $\bm{\Theta}$
		\State Define the error loss $\mathcal{L} = f(\hat{W}_{int}, W_{int})$ \Comment{Eq. (\ref{cost_L})}
		\State Build the optimizer $\mathcal{P} = f( \hat{\bm{\varepsilon}}, \hat{\bm{\sigma}},\Psi(\bm{\Theta}), \mathcal{L})$
\end{algorithmic}
\vspace{2px}
\textproc{Run training loop}
\begin{algorithmic}[1]
\State $epoch_i = 1$
\For {$epoch_i < \texttt{n}_{epoch}$}
    \State Compute the effective stress vector $\bar{\bm{\sigma}} = f(\bm{\Theta}, \hat{\bm{\varepsilon}})$ \Comment{Eq. (\ref{eq:effective_stress})}
    \State Decompose effective stress vector to $\bar{\bm{\sigma}}^{\pm} = f(\bar{\bm{\sigma}})$ \Comment{Eqs. (\ref{equ:effective_stress_composition_pos}) and (\ref{equ:effective_stress_composition_neg})}
    \State Compute equivalent stresses $\tau^{\pm} = f(\bm{\Theta}, \bar{\bm{\sigma}}^{\pm})$ \Comment{Eqs. (\ref{tau_plus}) and (\ref{tau_minus})}
    \State Calculate the damage thresholds $r^{\pm} = f(\bm{\Theta}, \tau^{\pm})$ \Comment{Eq. (\ref{scalar_damages})}
    \State Compute the damage variables $d^{\pm} = f(\bm{\Theta}, r^{\pm})$ \Comment{Eqs.  (\ref{damage_compression}) and (\ref{damage_tension})} 
    \State Compute the predicted stress $\bm{\sigma}_{pred} = f(\bar{\bm{\sigma}}, d^{\pm})$ \Comment{Eq. (\ref{eq:stress_tensor})}
       \State Compute the internal predicted work $W_{int}$
    \State Compute the error loss $\mathcal{L} = f(\hat{W}_{int}, W_{int})$ \Comment{Eq. (\ref{cost_L})}

    \If { $\left|\left|\Delta \boldsymbol{\Theta}\right|\right| < tol$ }
    \State end, return $\bm{\Theta}^\star$
    \Else 
    \State $epoch_i \to epoch_i + 1$
        \State $\mathcal{P}$\texttt{.optimize} \Comment{Scipy optimization procedure updates $\boldsymbol{\Theta}$}
    \EndIf
\EndFor
\end{algorithmic}
\end{algorithm}

\subsection{Virtual Laboratory} \label{sec:virtual_lab}
This research presents a technique that aims at modelling a highly heterogeneous and anisotropic material by means of homogenization procedures. For optimizing the material properties of the homogenized constitutive model, a wide variety of micro-model simulations have to be conducted in order to capture the heterogeneity of the masonry material. These micro-model simulations take into account the non-linear constitutive behaviors of the material’s components (brick and mortar) and its spatial distribution and orientation. Therefore, a Representative Volume Element (RVE) can be set up at the micro modeling scale in order to be exposed to a variety of boundary conditions.  The space where these analyses are carried out is called virtual laboratory (VL).

The subject of the VL is a micro model of a masonry wall. This RVE model must a) be able to represent both the homogeneous components of the masonry and b) consider the geometrical distribution and position of the components to each other - known as the masonry bond. Thus, a carefully designed geometrical model of the RVE has to be elaborated. At the same time, the composite components require being modeled by specific  constitutive laws (isotropic damage in this case detailed in Section \ref{sec:CL}) that are able to represent the individual component's behavior, respectively. 

\begin{figure}
    \centering
    \includegraphics[width=\linewidth]{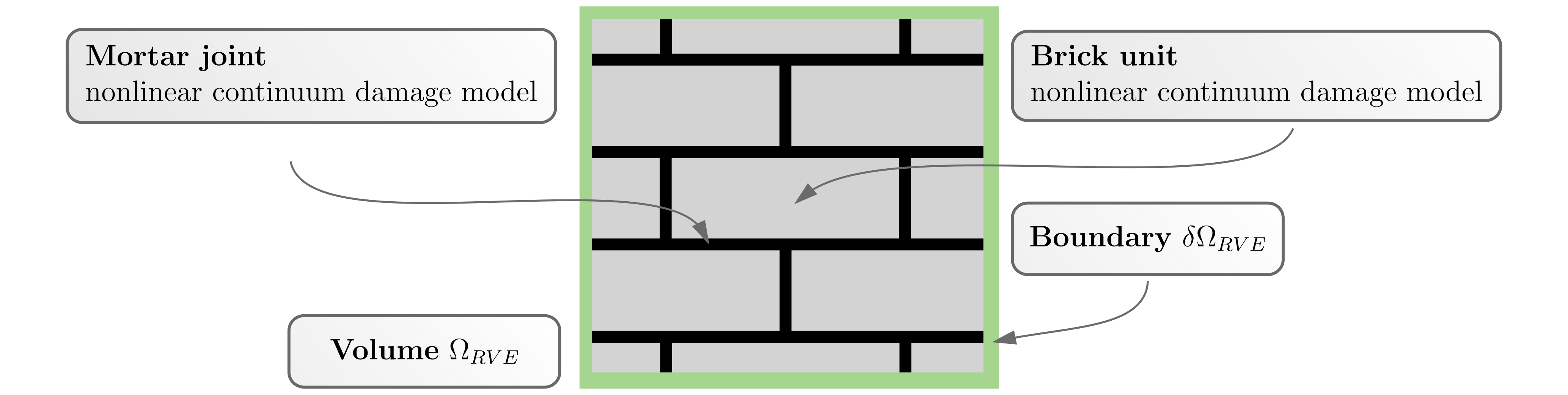}
    \caption{Example of micro-scale  RVE for the VL. The domain $\Omega_{RVE}$ consists of brick units and mortar joints, both numerically modeled as nonlinear homogeneous continuum materials. See also the RVE's boundary conditions $\delta\Omega_{RVE}$ \cite{PhilippThesis}.}
    \label{fig:machine_learning_masonry_rve_continuum_description}
\end{figure}

Fig. \ref{fig:machine_learning_masonry_rve_continuum_description} shows the schematic view of a micro model RVE of a masonry structure as it is considered in this research. Where $\Omega_{RVE}$ is the physical domain of the RVE and $\delta\Omega_{RVE}$ describes the boundary where the designed Dirichlet boundary conditions  are applied. Both homogeneous components, brick units and mortar joints, are modeled by nonlinear continuum damage models.

This paper adopts the concept of \textit{zero boundary displacement fluctuations} \cite{petracca2016a}. The boundary conditions are then applied as a displacement of the boundary $\delta\Omega_{RVE}$ by monotonically increasing its value at each analysis step. For the 2D analysis of the RVE the boundary conditions of a single virtual experiment are defined as follows:

\begin{align}
d_x = (\varepsilon_{xx} \cdot x + \varepsilon_{xy} \cdot y) \cdot t\label{equ:machine_learning_rve_boundary_disp_x}\\
d_y = (\varepsilon_{yy} \cdot y + \varepsilon_{xy} \cdot x) \cdot t \label{equ:machine_learning_rve_boundary_disp_y}
\end{align}

\noindent
where $d_x$ and $d_y$ are the displacements applied to all points of the RVE boundary $\delta\Omega_{RVE}$ in x- and y-direction, respectively. $t$ is the current time, so that $t\in [t_0, t_e]$, with $t_0$ as the start time and $t_e$ the end time of the VL analysis.  The scalar values $\varepsilon_{xx}$, $\varepsilon_{yy}$ and $\varepsilon_{xy}$ are the components of a previously defined strain vector in Voigt's notation; the strain vector then defines the imposed displacement field applied to the boundary of each  RVE. 

A determination of the boundary conditions in terms of a strain vector allows performing a large variety of virtual experiments. A modification of the strain vector in Eqs. \eqref{equ:machine_learning_rve_boundary_disp_x} and \eqref{equ:machine_learning_rve_boundary_disp_y} results in a specific boundary condition that deforms the RVE in a particular way. The virtually produced data should cover a wide range of possible deformations of the RVE to be the most representative as possible, leveraging the computational cost. However, since the calculations for generating the VL are performed in an offline stage, the computational cost is not that relevant.

Investigations made by Zaghi \textit{et al.} \cite{zaghi2018} show that all possible two dimensional strain states can be defined as the coordinates of a three dimensional sphere in a coordinate system where the components $\varepsilon_{xx}$, $\varepsilon_{yy}$ and $\varepsilon_{xy}$ of the strain vector are the axes. The strain vector can then be defined as follows

\begin{align}
\bm{\varepsilon} = \begin{pmatrix}
\varepsilon_{xx}\\\varepsilon_{yy}\\\varepsilon_{xy}
\end{pmatrix} = \lambda \begin{pmatrix}
 \cos \theta \\ \sin \theta \cos \phi \\ \sin \theta \sin \phi
\end{pmatrix}\label{equ:machine_learning_strain_vector_virtual_lab}
\end{align}

\noindent
where $\theta$, $\phi$ and $\lambda$ are the three parameters that define the coordinates of the sphere. The components of the strain vector depend on periodic sine and cosine functions. Thus, all possible strain configurations can be obtained by modifying the angles $\theta$ and $\phi$ in the interval $[-\pi, \pi]$. The scalar $\lambda$ is the norm of the strain vector so that

\begin{equation}
\lambda = || \bm{\varepsilon} || = \sqrt{\varepsilon_{xx}^2+\varepsilon_{yy}^2+\varepsilon_{xy}^2}.
\end{equation}

\begin{figure}[]
\centering
	\includegraphics[width=0.4\linewidth]{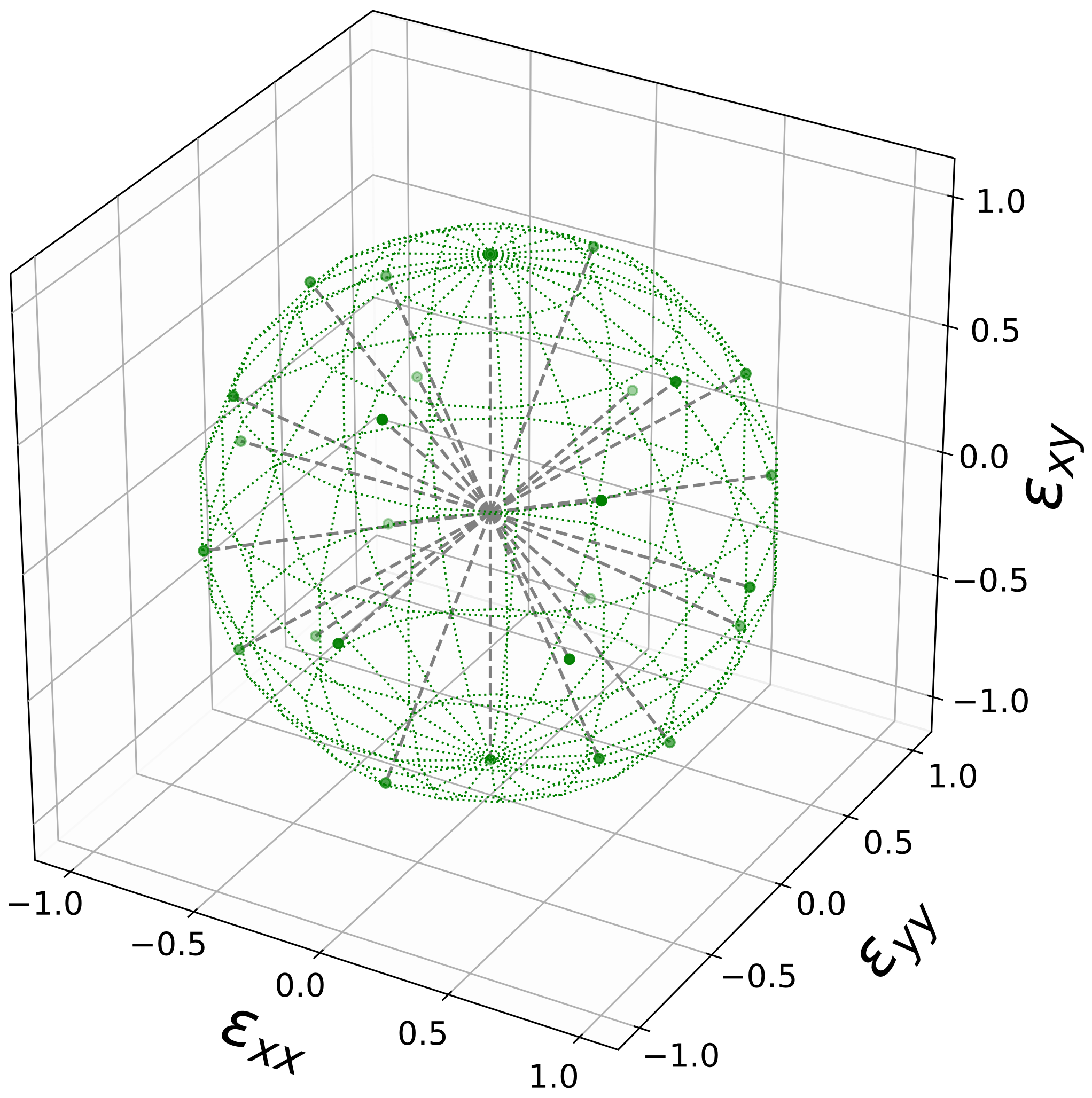} 
    \caption{Three dimensional strain space for the $26$ variation of the angles $\theta$ and $\phi$ in Eq. \eqref{equ:machine_learning_strain_vector_virtual_lab}.}
\label{fig:machine_learning_masonry_strain_sphere}
\end{figure}

\begin{figure}[]
\centering
	\includegraphics[width=0.7\linewidth]{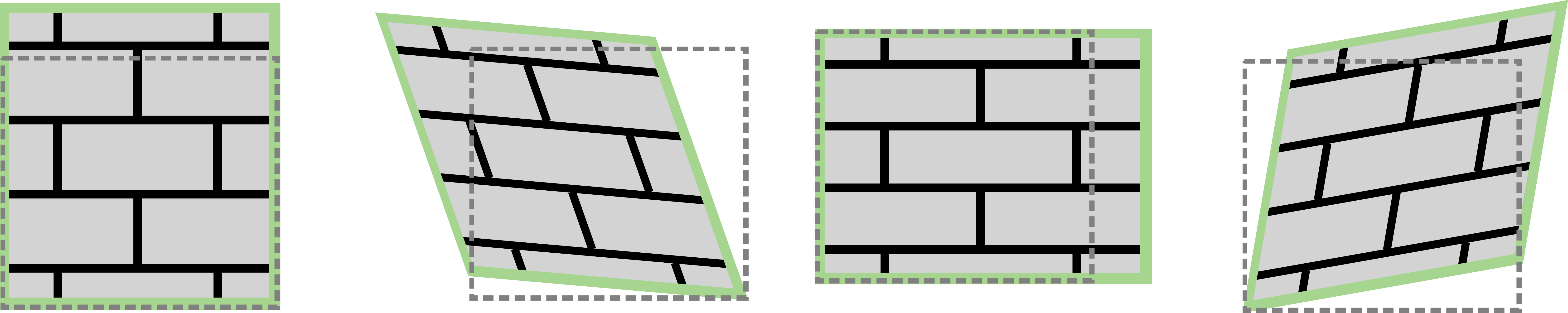} 
    \caption{Four deformed RVEs by applying a variation of boundary conditions}
\label{fig:machine_learning_masonry_four_rves_deformed}
\end{figure}

Fig. \ref{fig:machine_learning_masonry_strain_sphere} shows the three dimensional strain space and a variation of 26 combinations of the two angles in order to define the set of strain vectors. The norm of each strain vector is equal to 1. Fig. \ref{fig:machine_learning_masonry_four_rves_deformed} depicts the RVE exposed to four different boundary conditions caused by varying the strain vector of Eq. \eqref{equ:machine_learning_strain_vector_virtual_lab}. 

The solution of each boundary value problem of the virtual laboratory is calculated numerically via the FEM. A non-linear finite element analysis is performed for each virtual experiment, discretizing each material (brick and mortar) separately and using the most suitable constitutive law in each case. The number $n_{VL}$ of strain histories in the virtual laboratory depends on the number of variations of the strain vector $\bm{\varepsilon}$ of Eq. \eqref{equ:machine_learning_strain_vector_virtual_lab}, 26 in this work. Hence, for each $i$ strain history, all strain states are stored in a matrix \{$\boldsymbol{\varepsilon}_{h, i}$\} consisting of $n$ time steps strain row vectors (handled by numpy library) as follows:

\begin{equation}
\left\{ \boldsymbol{\varepsilon}_{h, i} \right\} = \begin{bmatrix}
\bm{\varepsilon}_{i, 0} & \bm{\varepsilon}_{i, 1} & \dots & \bm{\varepsilon}_{i, n}
\end{bmatrix}^T
\label{equ:machine_learning_strain_matrix_virtual_laboratory}
\end{equation}

Each finite element analysis does then take its corresponding entry of the matrix \{$\boldsymbol{\varepsilon}_{h, i}$\} in order to define the boundary conditions and runs until complete failure of the micro model can be achieved. 

For the finite element analyses, this research utilizes the open source framework \textsc{kratos multiphysics} \cite{kratos}. It offers the possibility to run the $n_{VL}$ analyses of the virtual laboratory by a user defined loop. So that all $n_{VL}$ analyses can be run automatically by a single command file. Moreover,  \textsc{kratos} includes the implementation and full linearization of the constitutive laws introduced in previous sections, so that the material components of the masonry RVE can be numerically modeled by an appropriate law for brittle materials. The modularity of \textsc{kratos} also allows to write user specific python files, that have access to the data structure while FE solving, \textit{e.g.} writing user specific output files automatically.

Being able to write such specific files is a great advantage for the preparation of the machine learning training data. The stresses exhibited during the analyses of the RVE must be up-scaled to a macro scale level in order to be used as training input for the prediction of a homogenized constitutive model. The following section describes the procedure of up-scaling the stresses. 

\paragraph{Up-scaling of strain and stress states}

The finite element solution provides the local deformations and stresses at all the Gauss points of the RVE. Hence, they must then be transformed to a macro level by up-scaling each entity in order to use the coupled strain and stress pairs as strain-internal work as input for the cost function $\mathcal{L}(\boldsymbol{\Theta})$ for training.

\begin{figure}[]
\centering
	\includegraphics[width=0.8\linewidth]{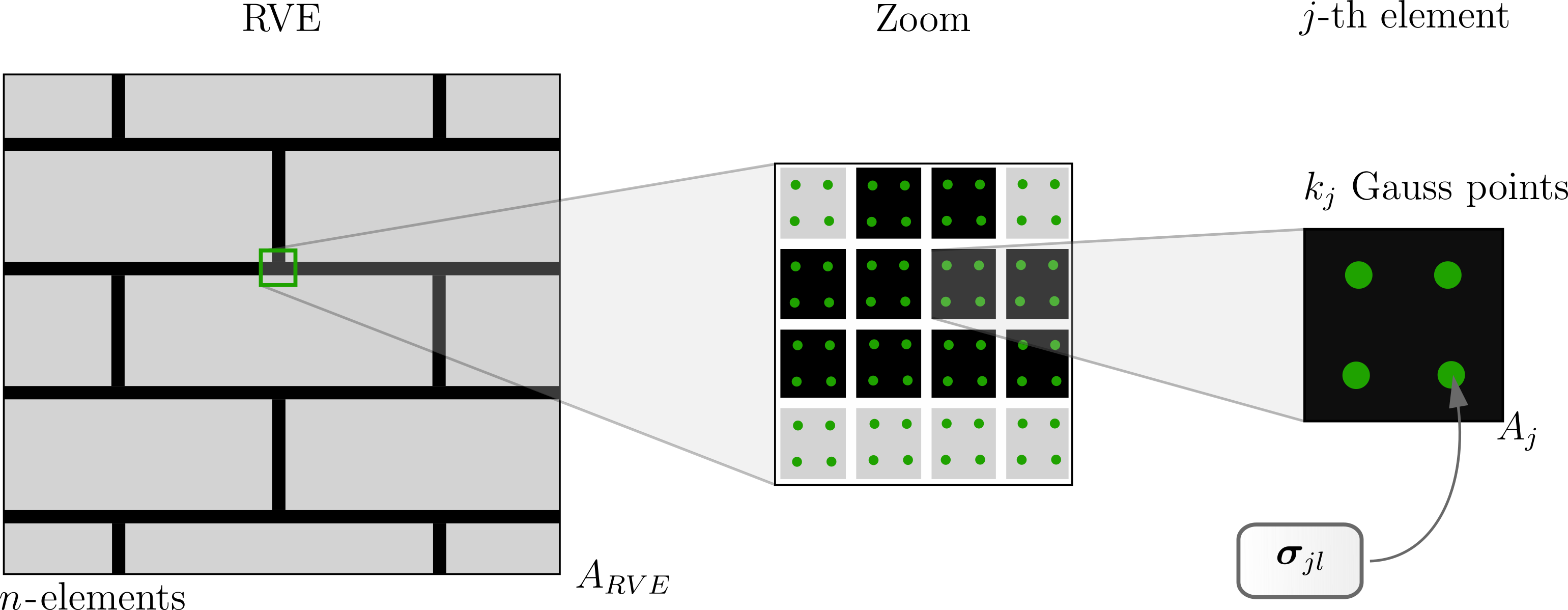} 
    \caption{Example showing the entities of Eq. \eqref{equ:machine_learning_mean_stress_upscaling} for the up-scaling procedure of a masonry RVE finite element model with 4 node quadrilateral elements each having $k_j=4$ Gauss integration points \cite{PhilippThesis}.}
\label{fig:ML_example_mean_stress_upscaling}
\end{figure}

This transition takes place by, for each strain state ($\bm{\varepsilon}_{i, t}$), computing a volume average value of all the IP  stresses of the RVE. Thus. The implemented procedure that computes the up-scaled stresses is defined as follows and is valid for 2D elements only:

\begin{equation}
\tilde{\bm{\sigma}}_{i, t} = \frac{1}{A_{RVE}}\sum_{j=1}^{n_{elem, RVE}}\Big[ \frac{A_j}{k_j} \sum_{l=1}^{k_j} \boldsymbol{\sigma}_{jl}\Big]
\label{equ:machine_learning_mean_stress_upscaling}
\end{equation}

\noindent
where $\tilde{\bm{\sigma}}_{i, t}$ is the  mean up-scaled stress vector of the history case $i$ at the time step $t$; $n_{elem, RVE}$ is the total number of finite elements of the RVE, $A_{RVE}$ is the total area of the RVE. $A_j$ is the area of the $j$-th element of the RVE, $k_j$ is the total number of Gauss points of the $i$-th element.  Fig. \ref{fig:ML_example_mean_stress_upscaling} shows the entities of Eq. \eqref{equ:machine_learning_mean_stress_upscaling} for a masonry RVE finite element model with four node quadrilateral elements each having $k_j = 4$ Gauss integration points. Once we have up-scaled the stresses of all time steps, we can build the VL stress matrix $\left\{ \tilde{\boldsymbol{\sigma}}_{h, i} \right\}$ for the $i$-th history as

\begin{equation}
\left\{ \tilde{\boldsymbol{\sigma}}_{h, i} \right\} = \begin{bmatrix}
\tilde{\boldsymbol{\sigma}}_{i, 0} & \tilde{\boldsymbol{\sigma}}_{i, 1} & \dots & \tilde{\boldsymbol{\sigma}}_{i, n}
\end{bmatrix}^T
\label{equ:machine_learning_stress_matrix_virtual_laboratory}
\end{equation}

This procedure of constructing a mean over the stresses and the strains to one single vector, results in values belonging to a representative single element (RSE). Thus this single element contains both the averaged vectors and is representative for the RVE.  It has the characteristic element size $l_{ch,RSE}$. The element size $l_{ch,RSE}$ is computed as the average value of all the characteristic element sizes of the RVE. 

The virtual laboratory is programmed as a loop over $n_{VL}$ analyses which depend on the parameters chosen to define the strain vector $\left\{ \boldsymbol{\varepsilon}_{h, i} \right\}$  of Eq. \eqref{equ:machine_learning_strain_matrix_virtual_laboratory}. Each numerical experiment of the virtual laboratory has a total analysis step number of $n$ by monotonically increasing the analysis time instance $t$. Alg. \ref{algorithm:machine_learning_virtual_laboratory} shows the numerical implementation of the automated loop of the virtual laboratory.

\begin{algorithm}[]
\caption{Automated procedure of the virtual laboratory}
\label{algorithm:machine_learning_virtual_laboratory}
\hspace*{\algorithmicindent} \textbf{input:} $n$, $n_{VL}$ \\
\hspace*{\algorithmicindent} \textbf{output:} $\tilde{\bm{\epsilon}}$, $\tilde{\bm{\sigma}}$\vspace{2px}\\
\textproc{START}
\begin{algorithmic}[1]
		\State Build the finite element model of the RVE 
		\State Set up the boundary strain vector $\left\{ {\boldsymbol{\varepsilon}}_{h, i} \right\}$ for all $i$ histories \Comment{Eq. (\ref{equ:machine_learning_strain_matrix_virtual_laboratory})} 
		\State Initialize the lists of the averaged stresses $\left\{ {\boldsymbol{\sigma}}_{h, i} \right\}$
		\State Initialize virtual laboratory loop $i = 1$
		\While {$i \leq n_{VL}$} 
		\State  Define the $i$-th boundary condition $d_{x/y} = f(\tilde{\boldsymbol{\varepsilon}}_{i, 0})$ \Comment{Eqs. (\ref{equ:machine_learning_rve_boundary_disp_x}), (\ref{equ:machine_learning_rve_boundary_disp_y})}
		\Procedure{\textsc{kratos} solving stage}{}
			\For {each analysis step $j \in n$}
			\State Apply boundary condition and solve the boundary value problem, incrementally
			\State Compute up-scaled stresses $\tilde{\bm{\sigma}}_j$ \Comment{Eq. (\ref{equ:machine_learning_mean_stress_upscaling}),} 
			\State Append upscaled entities: $\left\{ \tilde{\boldsymbol{\sigma}}_{h, i} \right\}$\texttt{.append}$(\tilde{\bm{\sigma}}_j)$ \Comment{Fill in Eq. \eqref{equ:machine_learning_stress_matrix_virtual_laboratory}}
			\EndFor
		\EndProcedure
		\State $i \to i+1$ \Comment{Go to next virtual experiment}
		\EndWhile
\end{algorithmic}
\textproc{END}
\end{algorithm}

\subsection{Data isotropization} \label{sec:isotropization}

The data produced in the VL serves as the training data of the machine learning model previously described in Sections \ref{sec:ML_introduction} and \ref{sec:ml_cl_model}. The model behind this technique is a constitutive nonlinear model based on linear elastic isotropy. It assumes an isotropic relation between the input strains $\bm{\epsilon}_{true}$ and the input stresses $\bm{\sigma}_{true}$. Thus the data coming from the VL should also follow an isotropic strain stress relationship.

This section shows that the raw data coming from the VL and the masonry RVE does not follow an isotropic elasticity relation.  Thus, a mapping procedure is presented that is able to isotropize the relation between strains and stresses. This method is called \textit{data isotropization}, which is basically a preparation of the input data for the machine learning model in order to ensure consistent isotropic input-output flow. All the necessary steps of the method are clarified in the following. An example that illustrates the problem at a linear elastic level can be found in \cite{PhilippThesis}. 

The method is based on a transformation between an anisotropic and an isotropic space. It requires a mapping matrix that can be computed from the raw elasticity matrix at the anisotropic level and the isotropic elasticity at the isotropic level. 

\subsubsection{Linear elastic properties} \label{subsec:data_iso_linear_elastic_props}

In order to compute both the raw anisotropic and the isotropic linear elastic matrices, the values of the strain vector $\tilde{\bm{\varepsilon}}$ and the stress vector $\tilde{\bm{\sigma}}$ must be extracted that are still in the linear elastic range. It is then satisfactory to gather the first entry of the $i$-th virtual experiment of all $n_{vl}$ analyses and store them in a matrix so that

\begin{equation}
\breve{\bm{\sigma}} = \bigg[ \tilde{\bm{\sigma}}_1 \hspace{5px} \tilde{\bm{\sigma}}_{1+n_1}\hspace{5px} \tilde{\bm{\sigma}}_{1+n_1 + n_2} \hspace{5px}\dots\hspace{5px} \tilde{\bm{\sigma}}_{1+n_1+n_2 + \dots + n_{n_{vl}-1}}\bigg]
\label{equ:machine_learning_extract_linear_stresses}
\end{equation}
\noindent
where $\breve{\bm{\sigma}}$ is a matrix of rank $3\times n_{vl}$ that includes the analysis results of the first analysis step of each virtual experiment extracted from the up-scaled stresses $\tilde{\bm{\sigma}}$. The subscript denotes the position of the first result value of each virtual experiment in the entire set of $\tilde{\bm{\sigma}}$; being $n_1, n_2, \dots, n_{vl}$  the scalar entries of a vector $\bm{n}$ that contains the number of analyses steps , $e.g.$ $n_1$ is the total number of analysis steps of the first virtual experiment, and so on. The vector $\bm{n}$  holds $n_{vl}$ entries. The computation of the corresponding strain vector $\breve{\bm{\epsilon}}$ containing only the first strains of each analysis can be constructed from $\tilde{\bm{\varepsilon}}$, accordingly. 

\paragraph{Raw elasticity matrix}
Given the  stresses  strains sets in the linear elastic range of all the virtual experiments, the raw elasticity matrix can be computed starting from linear elasticity as

\begin{equation}
\underbrace{\breve{\bm{\sigma}}}_{3\times n_{vl}} = \underbrace{\textbf{C}_{raw}}_{3\times 3} : \underbrace{\breve{\bm{\varepsilon}}}_{3\times n_{vl}}, \label{equ:machine_learning_elasticity_equation}
\end{equation}

\noindent
rearranging Eq. \eqref{equ:machine_learning_elasticity_equation} leads to the computation of the raw anisotropic elasticity matrix as a least squares problem

\begin{equation}
\textbf{C}_{raw} = \underbrace{\breve{\bm{\sigma}}}_{3\times n_{vl}} : \underbrace{\breve{\bm{\varepsilon}}^+}_{n_{vl}\times 3}
\label{equ:machine_learning_raw_elasticity_matrix}
\end{equation}

\noindent
where $\breve{\bm{\varepsilon}}^+$ denotes the \textit{Moore-Penrose} inverse of $\breve{\bm{\varepsilon}}$ that has a common use in solving least square problem (also known as \textit{pseudoinverse}). It is a generalization of the inverse matrix used for non symmetric matrices. So that $\bm{A}\bm{A}^+ = \bm{I}$ for a $(n \times k)$ matrix $\bm{A} (n \neq k)$.

\paragraph{Orthotropic elasticity matrix}

Masonry structures, when composed of  brick layers and  mortar joint  exhibit a different elastic (and also strength) response depending on the direction of study. The following application of the mapping procedure requires the orthotropic elasticity matrix of the masonry material. Therefore, a procedure is presented in order to transform the raw isotropic elasticity matrix to the orthotropic one. The computation of the Frobenius norm of the two matrices evaluates the procedure's error. The general orthotropic linear elastic matrix for two-dimensional problems in the plane stress condition reads as follows

\begin{equation}
\textbf{C} = \frac{1}{1-\nu_{12}\nu_{21}}\begin{bmatrix}
E_1 & \nu_{21}E_1 & 0\\
\nu_{12}E_2 & E_2 & 0\\
0& 0 & G
\end{bmatrix}
\label{equ:machine_learning_general_orthotropic_matrix_2d}
\end{equation}

\noindent
here, $E_{1/2}$ is the Young's modulus of the material in the 1- and the 2-direction, respectively, where such directions are perpendicular to each other. The same holds for the Poisson ratios $\nu_{12/21}$. $G$ is the shear modulus of the linear elastic material. The identification of the directions 1 and 2 is not required since the orthotropic matrix is utilized only for the computation of the closest isotropic matrix. The orthotropic linear elasticity matrix is symmetric and has zero entries in the first two entries of the third column and of the third row. These values must be equalized to zero, in order to obtain an orthotropic matrix, as 

\begin{equation}
C_{raw,ij} = 0 \text{ , with } i,j = (1,3),(2,3),(3,1),(3,2)
\label{equ:machine_learning_clean_raw_elasticity_matrix}
\end{equation}
\noindent
where $C_{raw,ij}$ is the entry of the raw elasticity matrix at the $i$-th row and the $j$-th column. Moreover, it must be ensured that the orthotropic elasticity matrix is symmetric. This can be ensured by

\begin{equation}
\textbf{C}_{ortho} = \frac{1}{2} \cdot \bigg(\textbf{C}_{raw} + \textbf{C}_{raw}^T \bigg)
\label{equ:machine_learning_orthotropic_elasticity_matrix}
\end{equation}

After having performed the transformation from the raw elasticity matrix to the orthotropic one, an error evaluation utilizing the Frobenius norm must be considered as in Kalkbrenner \cite{PhilippThesis}.

\paragraph{Closest isotropic elasticity matrix}
In this section, the closest isotropic-equivalent elastic matrix is computed from the previously calculated orthotropic elastic matrix $\textbf{C}_{ortho}$
In order to apply the previously mentioned transformation procedure between the orthotropic or and the isotropic scale, an isotropic elasticity matrix is required. Thus this paragraph applies the closed form definition for the computation of the closest isotropic matrix $\textbf{C}_{iso}$ to an anisotropic one (here $\textbf{C}_{raw}$), presented in Norris \cite{Norris06} and Rossi \textit{et al.} \cite{Rossi21}. The procedure is defined as follows

\begin{equation}
\textbf{C}_{iso} = 3 \Big( \frac{\alpha^{\star}}{3}\kappa^{\star}\Big) \textbf{J} + 2 \mu^{\star} \textbf{K}
\label{equ:machine_learning_closest_iso_matrix_closed_form}
\end{equation}
\noindent
where the coefficient $\alpha^{\star} = 2$, the matrix $\textbf{J} = \textbf{tt}^T$, with $\textbf{t}^T = (2^{-0.5},2^{-0.5},0)$, the matrix $\textbf{K} = \textbf{I}_4-\textbf{J}$. Where $\textbf{I}_4 = diag\big(1,1,0\big)$. The values $\kappa^{\star}$ and $\mu^{\star}$ depend on the original anisotropic matrix in the following way

\begin{equation}
\mu^{\star} = 0.2 ( c_{00} - 2 c_{01} +c_{11} +c_{22} )\\
\label{equ:machine_learning_closest_iso_matrix_closed_form_mu}
\end{equation}

\begin{equation}
\kappa^{\star} =\frac{\textbf{m}^T \textbf{C}_{aniso} \textbf{m}}{\alpha^2}
\label{equ:machine_learning_closest_iso_matrix_closed_form_kappa}
\end{equation}
\noindent
being $\textbf{m}= \big(1,1,0\big)^T$ and the $c_{ij}$ define the corresponding entry of the anisotropic matrix $\textbf{C}_{raw}$. 
The definition of the closest isotropic matrix is based on the Frobenius norm $|| \textbf{C}_{raw} - \textbf{C}_{iso}||_F$.  

\subsubsection{Isotropic mapping of up-scaled stresses} \label{subsec:data_iso_mapping}

In this work, a classical transformation by means of mapping matrix  \textbf{T} has been used to define the bijective function between the orthotropic and isotropic spaces \cite{oller_anisot_1, oller_anisot_2, Cornejo_thesis, Pel2011ContinuumDM, PELA2013957, PELA2014196}. This relationship is

\begin{equation}
\textbf{C}_{ortho} = \textbf{T}^T :\textbf{C}_{iso} : \textbf{T}
\label{equ:machine_learning_ortho_iso_transformation_general}
\end{equation}
\noindent
in which  $\textbf{T}$ is a transformation matrix that still needs to be defined. In order to do so, let the matrix square roots $\sqrt{\textbf{C}_{ortho}}$ and $\sqrt{\textbf{C}_{iso}}$ of both the matrices $\textbf{C}_{ortho}$ and $\textbf{C}_{iso}$ be considered. Such matrix square roots do exist, since $\textbf{C}_{ortho}$ and $\textbf{C}_{iso}$ are symmetric positive definite (SPD). Then, Eq. \eqref{equ:machine_learning_ortho_iso_transformation_general}
 can be rearranged by utilizing the symmetric square root matrices as follows
 
\begin{equation}
\sqrt{\textbf{C}_{ortho}} : \sqrt{\textbf{C}_{ortho}} = \textbf{T}^T :  \sqrt{\textbf{C}_{iso}} : \sqrt{\textbf{C}_{iso}} : \textbf{T}
\end{equation} \noindent
from which one can compute the transformation matrix according to

\begin{equation}
\textbf{T} = \Big(\sqrt{\textbf{C}_{iso}}\Big)^{-1} : \sqrt{\textbf{C}_{ortho}}
\label{equ:machine_learning_transformation_matrix}
\end{equation}

 This mapping strains and stresses to a fictional isotropic space must be applied to the entire set of up-scaled strains and stresses coming from the virtual laboratory. Then the isotropic up-scaled stresses and strains read
  
\begin{align}
\tilde{\bm{\varepsilon}}_{iso} &= \textbf{T} : \tilde{\bm{\varepsilon}}\label{equ:machine_learning_iso_strains_mapping}\\
\tilde{\bm{\sigma}}_{iso} &= \textbf{T}^{-T} : \tilde{\bm{\sigma}}\label{equ:machine_learning_iso_stresses_mapping}
\end{align}

\noindent where $\tilde{\bm{\varepsilon}}$ and $\tilde{\bm{\sigma}}$ are all the coupled strains and stresses coming from the virtual laboratory.

\subsection{Post machine learning constitutive model} \label{sec:post_ml_cl}

This section presents the procedure and the implementation that enables to use the result of the machine learning homogenization technique to properly train a macro-scale constitutive model. 

\begin{figure}[]
\centering
	\includegraphics[width=0.8\textwidth]{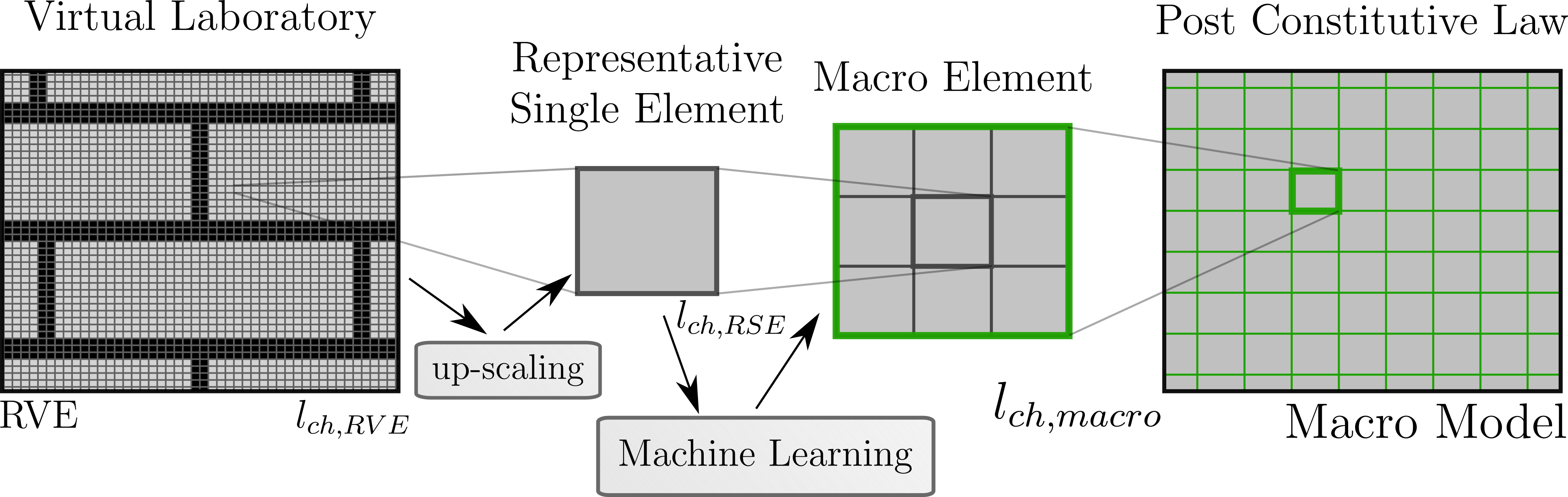} 
    \caption{Chart of the operational flow in order to obtain the post-machine-learning constitutive model. Showing the fictitious position of the representative single element in a macro model \cite{PhilippThesis}.}
\label{fig:machine_learning_general_operation_flow}
\end{figure}

\noindent
Fig. \ref{fig:machine_learning_general_operation_flow} shows a flowchart of how the post-machine-learning constitutive law is extracted from the machine learning procedure by considering all analyses at the multiple scales. In order to obtain a fully consistent post machine learning constitutive law, several issues must be tackled:  the analysis of the virtual laboratory and the subsequent  up-scaled procedure that averages the stresses and strains of the RVE in a representative single element (RSE). These stresses are then sent to the machine learning technique and train a constitutive model, that represents a homogeneous damage constitutive model for the RSE. The RSE is then multiple times fictitiously positioned in the element of a homogeneous macro model. In this way the constitutive model trained at the RSE level can be reflected and utilized for the analysis at macro scale. A factor $\omega_{ch}$ is introduced, that counts the number of multiple positions of the RSE in the macro element. It is important for the energy regularization and mesh independent results \cite{PhilippThesis} during the homogenization technique and it is defined as follows

\begin{align}
\omega_{ch} = \frac{l_{ch,macro}}{l_{ch,RSE}}
\label{equ:machine_learning_post_ml_omega}
\end{align}

\noindent where $l_{ch,RSE}$ is the mean value of  the characteristic lengths $l_{RVE}$ of the elements of the RVE and $l_{ch,macro}$ is the characteristic length of the element at the macro scale.

The parameters of the post-machine-learning constitutive law have been obtained during the machine learning procedure and are stored as optimized parameters in $\bm{\Theta}^{\star}$. Most of the parameters included in $\bm{\Theta}^{\star}$ can be directly used for the post-machine-learning constitutive law. In order to obtain size independent results when using a certain amount of fracture energy, an energy regularization is required.

\paragraph{Energy regularization} \label{subsec:energy_regularization}

As shown in Fig. \ref{fig:machine_learning_general_operation_flow}, two different scales of mesh sizes are present. On the one hand there is the characteristic length $l_{ch,RSE}$ of the RSE and on the other there is the  characteristic length of the element at the macro scale.
\noindent
The same problem occurs for two different micro models with different finite element meshes. Thus, an energy regularization must be performed in order to obtain mesh and size independent results. For both the damage evolution laws present in this paper, namely the exponential softening and the B\'ezier like hardening softening behavior, the energy is regularized over the element's size. This procedure has been introduced in Section \ref{sec:damage}. It is simple and straight forward, since the energy regularization is based on the user input of the fracture energy $G^-$ or $G^+$. Both values are divided by the characteristic element length $l_{ch}$ in order to compute a  fracture energy density $g^-$ or $g^+$. These specific values then regularize the energy over the element. The  energy regularization for classical applications (Section \ref{sec:CL}) is done via

\begin{equation}
G^{\pm} \longrightarrow g^{\pm} = \frac{G^{\pm}}{l_{ch}}
\label{equ:machine_learning_energy_reg_flow_general_procedure}
\end{equation}

However, this operation is not necessary for the post machine learning constitutive law, since the specific fracture energies $g^{\pm}$ can be directly computed from the optimized constitutive law parameters $\bm{\Theta}^{\star}$ and the known characteristic element length $l_{ch,RSE}$ of the RSE. However, it has been shown in Fig. \ref{fig:machine_learning_general_operation_flow} that the element sizes of the RSE and the macro model are unequal. Thus it must be shown that the fracture energy $G_{c/t}$ of the post machine learning constitutive law is computed in a consistent way. It all depends on the elemental size. The difference in elemental size of the both scales is measured by the factor $\omega_{ch}$ as introduced in Eq. \eqref{equ:machine_learning_post_ml_omega}. The calculations introduced in \cite{PhilippThesis} proof the proposed approach.

\paragraph{Space transformation} \label{subsec:space_transformation}

The constitutive model subjected in this work (Section \ref{sec:CL}) and implemented to the machine learning technique is based on an isotropic strain stress relation as shown in Eq. \eqref{eq:effective_stress}. That is why the strains and stresses, coming from the virtual laboratory and contributing as inputs of the machine learning model, have been transformed into an isotropic relation, as introduced in Sec. \ref{sec:isotropization}.
Without transforming, their real relation, would be an orthotopic or even an anisotropic one and would not serve for the constitutive model implemented into the machine learning procedure of this paper. 

This mapping procedure of transforming between multiple spaces (anisotropic - isotropic - anisotropic) must also be included to the post machine learning constitutive model for the finite element analysis of macro scale masonry structures.
 
The input strains $\tilde{\bm{\epsilon}}$ and the predicted stresses $\tilde{\bm{\sigma}}_{pred}$ coming from the machine learning model are defined in an isotropic space. Hence, the trained constitutive law is not directly applicable to the macro modeling approach  of masonry, since it does not include the anisotropy of masonry. Thus, during the macro scale analysis, the strains coming from the finite element solving stage are defined in the anisotropic space and must be transformed to an isotropic space, in order to apply the optimized constitutive law. The strain state at the isotropic space can then be obtained by applying the mapping procedure as follows

\begin{align}
\tilde{\bm{\epsilon}}_{iso} = \textbf{T} : \tilde{\bm{\varepsilon}}
\label{equ:machine_learning_transformation_post_epsilon}
\end{align}

Where $\tilde{\bm{\epsilon}}$ is the anisotropic strain state coming from the finite element solving stage. The transformation matrix \textbf{T} has been obtained previously, while transforming the input training data for the optimization procedure at machine learning level. $\tilde{\bm{\epsilon}}_{iso}$ is the strain state at the isotropic space and is ready to run through the procedure described in Algorithm \ref{algorithm:machine_learning_optimization_procedure}. This results in $\tilde{\bm{\sigma}}_{iso}$, the stress state at the isotropic space. In order to finalize the post machine learning constitutive law, the stress state at isotropic level must be transformed back to the initial space by applying a transformation as follows

\begin{align}
\tilde{\bm{\sigma}} = \textbf{T}^T : \tilde{\bm{\sigma}}_{iso}
\label{equ:machine_learning_transformation_post_sigma}
\end{align}

\begin{figure}[]
\centering
	\includegraphics[width=0.8\linewidth]{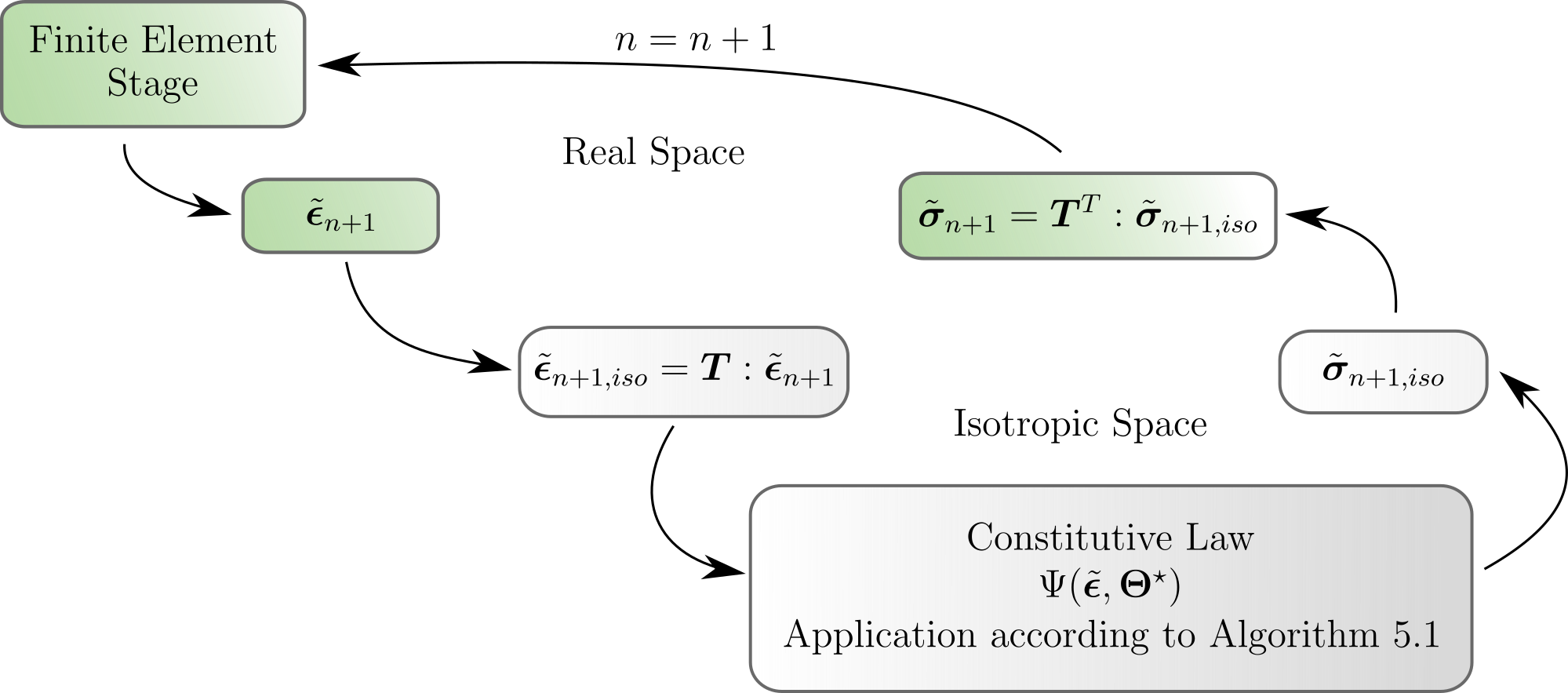} 
    \caption{Overview of general application procedure of the post-machine-learning constitutive law for the analysis of structures at the macro scale \cite{PhilippThesis}.}
\label{fig:machine_learning_overview_post_ml_cl}
\end{figure}

Where $\tilde{\bm{\sigma}}$ is the computed stress at the anisotropic space obtained by applying the post-machine-learning constitutive model. Fig. \ref{fig:machine_learning_overview_post_ml_cl} shows an overview of the mapping procedure and its implementation to the constitutive law at an isotropic space.  

\section{Numerical example - Diagonal compression test of a Flemish bond masonry wall} \label{sec:example}

This section presents an application example of the propsed machine learning homogenization technique for masonry structures. It demonstrates that a unique constitutive law can be found by means of machine learning that is able to represent the properties of a masonry high fidelity model.

\subsection{High fidelity micro scale virtual laboratory campaign} \label{sec:application_virtual_lab}

The representative volume element is based on the analyses made in Kalkbrenner \cite{PhilippThesis}, here revisited with different material properties. This section treats the calibration of the micro model properties for the diagonal compression test of a masonry wall. Table \ref{table:machine_learning_law_parameters_diagonal_compression} shows the material properties of the components according to Petracca \textit{et al.} \cite{Petracca2017}.

\begin{table}[h]
\caption{Material properties, brick unit and mortar joint, for the numerical analysis of the shear compression test applied to a Flemish bond masonry wall and considered for the presented virtual laboratory.}
\vspace{-5px}
	\begin{center}
		\begin{tabular}{cccccccccccccc}
\multicolumn{11}{c}{Brick unit}\\
 $E$ & $\nu$ & $f_{p}^+$ & $G^+$ & $f_{0}^-$ & $f_{p}^-$ & $f_{r}^-$ & $\epsilon_p^-$ & $G^-$ &$k_b$ & $\kappa$ & $c_1$ & $c_2$ & $c_3$\\
 \hline
 $7.0$ & $0.2$ & $2.0$ & $0.08$ & $8.0$ & $12.0$ & $1.0$ & $0.004$ & $6.0$ & $1.2$ & $0.0$ & 0.65 & 0.5 & 1.5\\
 \hline
$[GPa]$ & $-$ & $[MPa]$ & $[\frac{N}{mm}]$ & $[MPa]$ & $[MPa]$ & $[MPa]$ & $-$ & $[\frac{N}{mm}]$ & $-$ & $-$ & $-$ & $-$ & $-$\\

 & & & & & & & & & & \\

\multicolumn{11}{c}{Mortar joint}\\
 $E$ & $\nu$ & $f_{p}^+$ & $G^+$ & $f_{0}^-$ & $f_{p}^-$ & $f_{r}^-$ & $\epsilon_p^-$ & $G^-$ & $k_b$ & $\kappa$ & $c_1$ & $c_2$ & $c_3$\\
 \hline 
 $1.8$ & $0.2$ & $0.12$ & $0.016$ & $3.0$ & $10.0$ & $2.0$ & $0.04$ & $80.0$ & $1.2$ & $0.16$ & 0.65 & 0.5 & 1.5\\
 \hline
$[GPa]$ & $-$ & $[MPa]$ & $[\frac{N}{mm}]$ & $[MPa]$ & $[MPa]$ & $[MPa]$ & $-$ & $[\frac{N}{mm}]$ & $-$ & $-$ & $-$ & $-$ & $-$\\
		\end{tabular}
		\label{table:machine_learning_law_parameters_diagonal_compression}
	\end{center}
\vspace{-10px}
\end{table}

\begin{figure}[h]
    \centering
    \begin{subfigure}[b]{0.4\textwidth}
        \includegraphics[width=\textwidth]{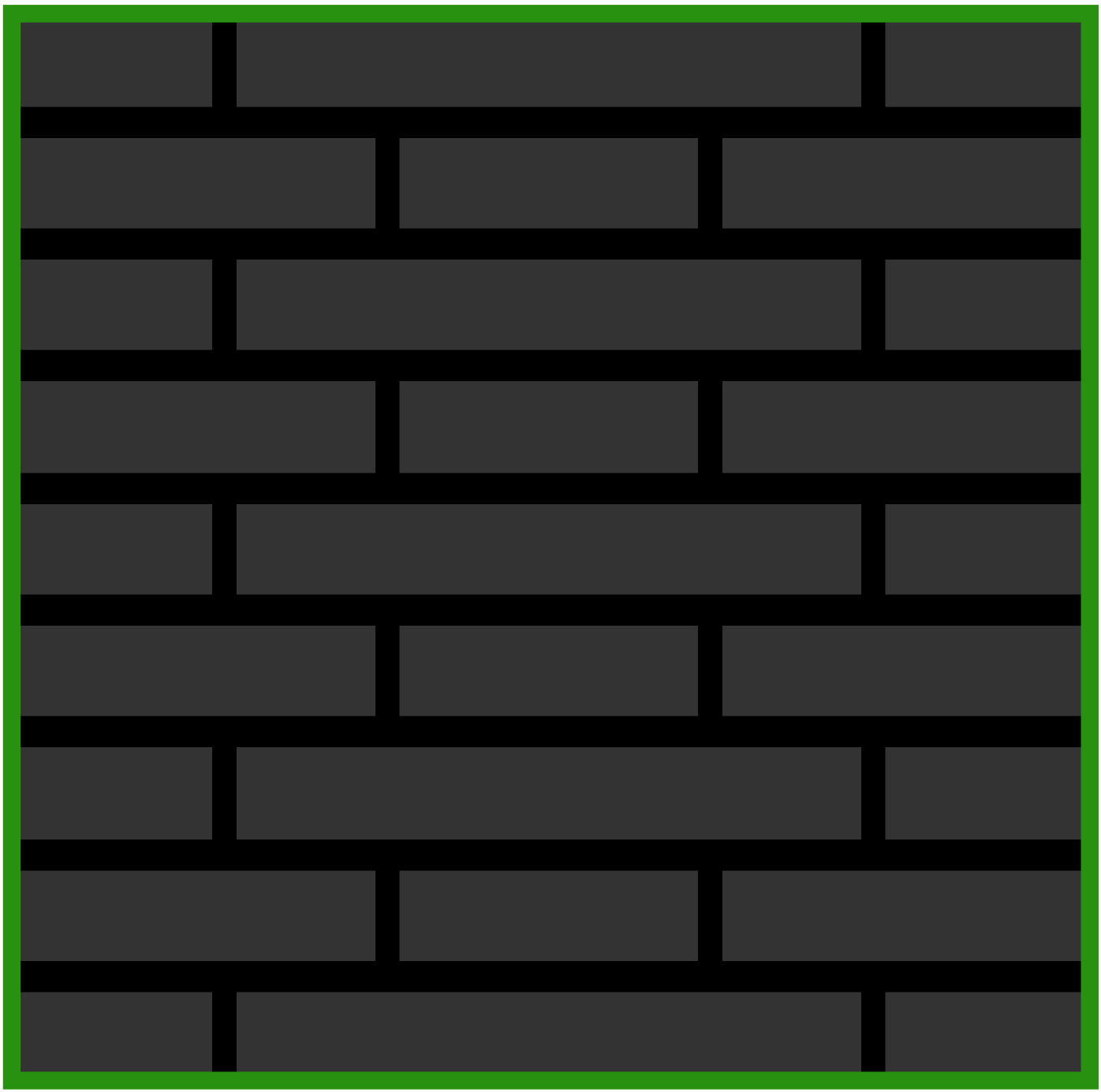}
        \caption{RVE micro model}
        \label{fig:machine_learning_application_rve}
    \end{subfigure}
    \begin{subfigure}[b]{0.4\textwidth}             
        \includegraphics[width=\textwidth]{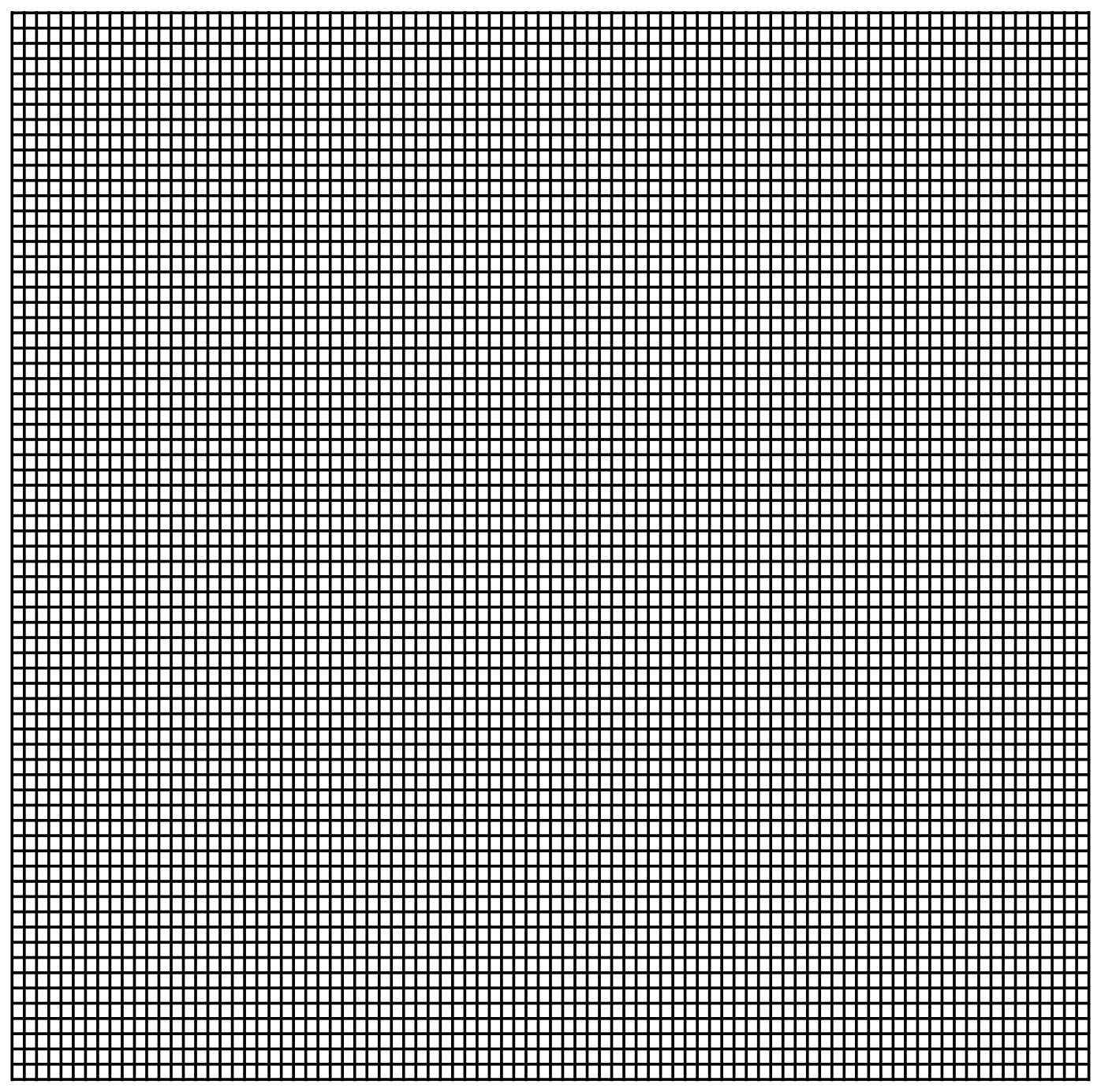}
        \caption{FE mesh, 6319 nodes and 6160 linear quads.}
        \label{fig:machine_learning_application_rve_mesh}
    \end{subfigure}\vspace{-5px}
    \caption{Virtual laboratory: RVE of the micro scale finite element analysis, showing a) the micro model and b) the finite element mesh discretization \cite{PhilippThesis}.}\label{fig:machine_learning_application_rve_and_mesh}
\end{figure}

Fig. \ref{fig:machine_learning_application_rve_and_mesh} shows the numerical model of the RVE considered for this virtual laboratory, identical to the one used in Kalbrenner \cite{PhilippThesis}. It depicts the mortar and the brick elements, respectively, and their finite element mesh discretization.

Both materials, either the mortar joints or the brick units, follow an isotropic $d^+/d^-$ damage law, where tension follows an exponential softening, and compression follows the exposed B\'ezier  hardening-softening behavior as introduced in Section \ref{sec:CL}

The virtual laboratory campaign consists of $n_{VL} = 26$ numerical analyses. Each analysis imposes a different strain field at the boundaries of the RVE. The analyses are performed until complete failure of the RVE is obtained.  The results of the virtual experiments are discussed and stored according to the up-scaling procedure introduced in Section \ref{sec:virtual_lab}. Table \ref{table:machine_learning_application_summary_cases} shows a summary of the results obtained in each case. It includes the unit vector of the boundary applied strain state $\tilde{\bm{\varepsilon}}$ that demonstrates the multi-directional deformations applied on the masonry RVE. Column 3 shows the stress state induced by the applied boundary conditions. While C stands for compression and T for tension. In total, there are $5$ compression/compression, $5$ tension/tension and $16$ mixed tension/compression states. Since the strength in tension is lower than the one in in compression for both the brick units and the mortar joints, the models with tension/compression states fail mainly due to tension stresses.

\begin{table}[]
\caption{Virtual laboratory: summary showing the unit vectors of the applied boundary strains in engineering notation, the corresponding stress state (C = Compression, T = Tension), and the indication whether damage in tension or compression is significant}
\vspace{-5px}
	\begin{center}
		\begin{tabular}{c|ccc}
\hline
\multirow{2}{*}{Case} & Unit boundary strain & Stress state & \multirow{2}{*}{Damage in} \\
 & ($\varepsilon_{xx}$, $\varepsilon_{yy}$, $\gamma_{xy}$) & ($\sigma_1$/$\sigma_2$) & \\
\hline
$1$ & ($-1.0$, $0.0$, $0.0$) & C/C & C \\
$2$ & ($-0.71$, $-0.71$, $0.0$) & C/C & C \\
$3$ & ($-0.53$, $-0.38$, $-0.76$) & C/C & C \\
$4$ & ($-0.53$, $-0.38$, $0.76$) & C/C & C \\
$5$ & ($-0.45$, $0.0$, $-0.89$) & T/C & T \\
$6$ & ($-0.45$, $0.0$, $0.89$) & T/C & T \\
$7$ & ($-0.53$, $0.38$, $-0.76$) & T/C & T \\
$8$ & ($-0.53$, $0.38$, $0.76$) & T/C & T \\
$9$ & ($-0.71$, $0.71$, $0.0$) & T/C & T \\
$10$ & ($0.0$, $-1.0$, $0.0$) & C/C & C \\
$11$ & ($0.0$, $-0.45$, $-0.89$) & T/C & T \\
$12$ & ($0.0$, $-0.45$, $0.89$) & T/C & T \\
$13$ & ($0.0$, $0.0$, $-1.0$) & T/C & T \\
$14$ & ($0.0$, $0.0$, $1.0$) & T/C & T \\
$15$ & ($0.0$, $0.45$, $-0.89$) & T/C & T \\
$16$ & ($0.0$, $0.45$, $0.89$) & T/C & T \\
$17$ & ($0.0$, $1.0$, $0.0$) & T/T & T \\
$18$ & ($0.71$, $-0.71$, $0.0$) & T/C & T \\
$19$ & ($0.53$, $-0.38$, $-0.76$) & T/C & T \\
$20$ & ($0.53$, $-0.38$, $0.76$) & T/C & T \\
$21$ & ($0.45$, $0.0$, $-0.89$) & T/C & T \\
$22$ & ($0.45$, $0.0$, $0.89$) & T/C & T \\
$23$ & ($0.53$, $0.38$, $-0.76$) & T/T & T \\
$24$ & ($0.53$, $0.38$, $0.76$) & T/T & T \\
$25$ & ($0.71$, $0.71$, $0.0$) & T/T & T \\
$26$ & ($1.0$, $0.0$, $0.0$) & T/T & T \\
\hline
		\end{tabular}
		\label{table:machine_learning_application_summary_cases}
	\end{center}
\vspace{-10px}
\end{table}

\begin{figure}[]
\centering
	\includegraphics[width=0.9\linewidth]{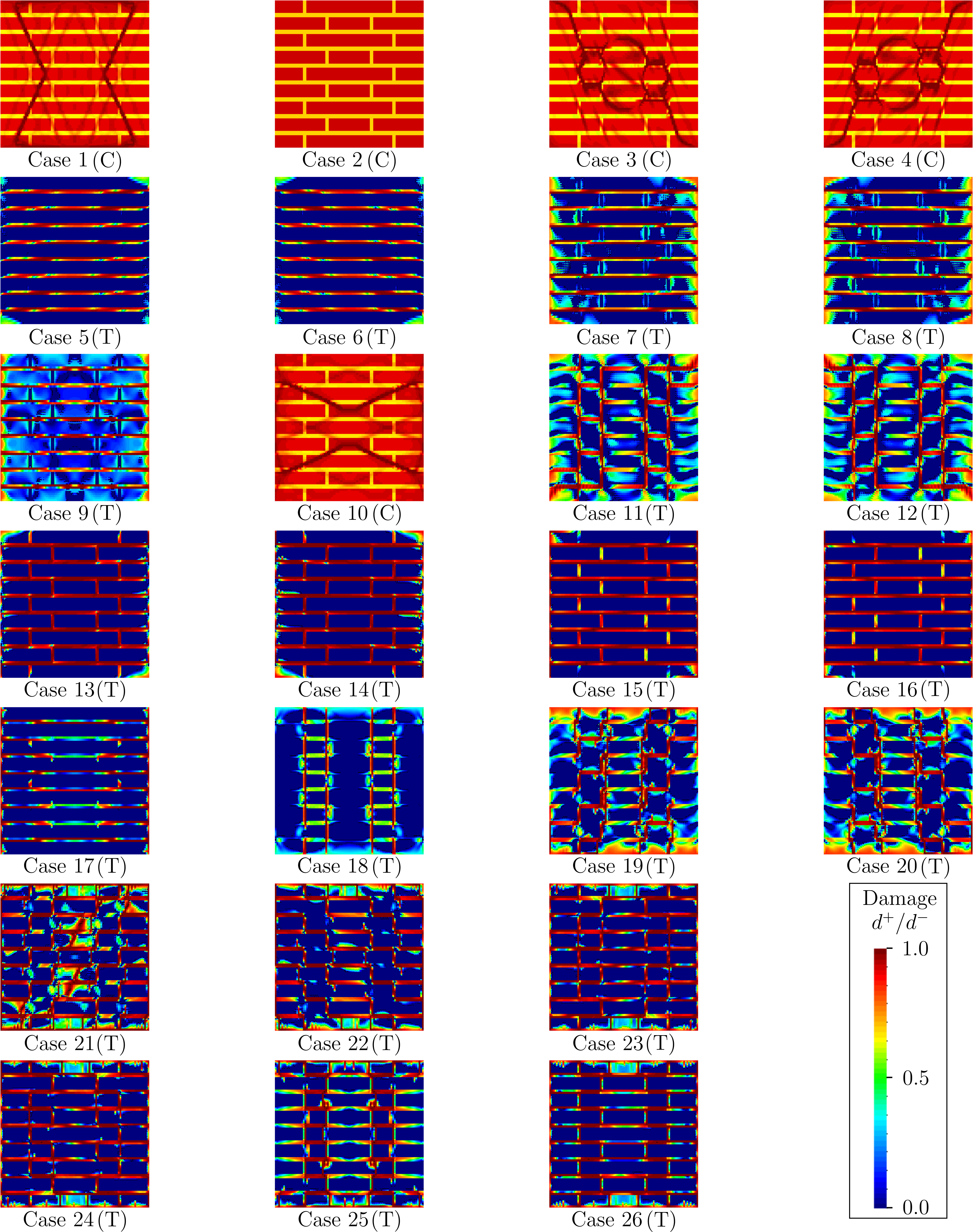} 
    \caption{Flemish bond RVE virtual laboratory: damage contour plots of the $26$ cases considered in the virtual laboratory, showing the damage variables $d^+/d^-$ depending on the significant damage (T for tension, C for compression) \cite{PhilippThesis}}
\label{fig:machine_learning_application_vl_damage_plots}
\end{figure}

Fig. \ref{fig:machine_learning_application_vl_damage_plots} shows an overview of the damage patterns at an ultimate state of each case analyzed in the virtual laboratory. It demonstrates the variety of different crack patterns and failure mechanisms. The patterns of each case are indicated whether the damage variable $d^+$ or $d^-$ leads to failure of the RVE masonry wall.

The virtual laboratory is finalized and serves as the data production part of the machine learning homogenization technique. The outcomes are the coupled sets of strains $\tilde{\bm{\varepsilon}}$ and stresses $\tilde{\bm{\sigma}}$ of each analysis step for all the considered virtual experiments. These data require further preparation as shown in the following.

The following step is the preparation and isotropization of the training data. Therefore the procedure described in \ref{sec:isotropization} is applied. The following orthotropic (Eq. \eqref{equ:machine_learning_ortho_iso_transformation_general}), isotropic (Eq. \eqref{equ:machine_learning_application_iso_elasticity_matrix_values}) and transformation (Eq. \eqref{equ:machine_learning_application_transformation_matrix_values}) matrices can be obtained. 

\begin{equation}
\textbf{C}_{ortho} = \begin{bmatrix}
5.442 & 0.83 & 4.99 \cdot 10^{-7}\\
0.83 & 4.291 & 6.70 \cdot 10^{-7}\\
4.99 \cdot 10^{-7} & 6.70 \cdot 10^{-7}  \cdot 10^{-6} & 1.707
\end{bmatrix} GPa
\label{equ:machine_learning_application_raw_elasticity_matrix_values}
\end{equation}

\begin{equation}
\textbf{C}_{iso} = \begin{bmatrix}
4.686 & 1.014 & 0.0\\
1.014 & 4.686 & 0.0\\
0.0 & 0.0 & 1.836
\end{bmatrix} GPa
\label{equ:machine_learning_application_iso_elasticity_matrix_values}
\end{equation}

\begin{equation}
\textbf{T} = \begin{bmatrix}
1.084 \cdot 10^0 & -1.686 \cdot 10^{-2} & 5.008 \cdot 10^{-8} \\
-3.036 \cdot 10^{-2} & 9.604 \cdot 10^{-1} & 8.359 \cdot 10^{-8} \\
 9.407\cdot 10^{-8} & 1.415 \cdot 10^{-7} & 9.641 \cdot 10^{-1}
\end{bmatrix}
\label{equ:machine_learning_application_transformation_matrix_values}
\end{equation}

\subsection{Machine learning homogenization} \label{sec:application_ml_technique}

Once the isotropic elastic properties are known and its corresponding mapper operator between the isotropic and the orthotropic spaces have been estimated (Eq. \eqref{equ:machine_learning_application_transformation_matrix_values}), a parameter optimization of the non-linear part of the calculation is performed. The macro model to be optimized is the same one we defined in Section \ref{sec:CL}, this means that we need to automatically estimate a set of 12 (we discounted the elastic properties) parameters $\boldsymbol{\Theta}$, \textit{i.e.}:

\begin{equation}
\setcounter{MaxMatrixCols}{20}
    \boldsymbol{\Theta} = 
    \begin{bmatrix}
        f_0^+ & G_f^+ & k_b & \kappa & G_f^- & f_p^- & \varepsilon_p^- & f_0^- & f_r^- & c_1 & c_2 & c_3
    \end{bmatrix}
\end{equation}

\noindent
where $f_0^+$ is the tensile yield stress, $G_f^+$ the tensile fracture energy, $k_b$ the biaxial compression multiplier, $\kappa$ the shear compression reductor coefficient, $G_f^-$ the compression fracture energy, $f_p^-$ the peak compression yield stress, $\varepsilon_p^-$ the peak yield strain, $f_0^-$ the compression yield stress, $f_r^-$ the residual compression stress and $c_1, c_2, c_3$ the three Bézier curve parameters. This parameter set is provided to the optimizer together with a user defined initial guess $\boldsymbol{\Theta}_0$ based on experience as (IS units, separated in two rows only for visualization):

\begin{equation}
\setcounter{MaxMatrixCols}{20}
    \boldsymbol{\Theta}_0 = 
    \begin{bmatrix}
        3.5\cdot10^5 & 500 & 1.15 & 2.7443\cdot10^{-5} & 7.781\cdot10^2 & 10^7 \\
        6.10\cdot10^{-3} & 3.997\cdot10^6 & 10^4 & 0.49547 & 0.6 & 2.1997
    \end{bmatrix}
\end{equation}

\noindent together with reasonable bounds $\left\{ \boldsymbol{\Theta} \right\}$ for each parameter to help the convergence of the optimizer to physically sound solutions. The used bounds are given in Table \ref{tab:my_bounds}.

\begin{table}[htbp]
    \centering
    \begin{tabular}{|l|l|l|}
        \hline

        Parameter & Lower Bound & Upper Bound \\
        \hline
        $f_0^+$ & $1.5 \cdot 10^5$ & $5.0 \cdot 10^5$ \\
        $G_f^+$ & 100 & $2.0 \cdot 10^3$ \\
        $k_b$ & 1.15 & 1.7 \\
        $\kappa$ & 0.0 & 0.2 \\
        $G_f^-$ & 600 & 1500 \\
        $f_p^-$ & $7.0 \cdot 10^6$ & $11.0 \cdot 10^6$ \\
        $\varepsilon_p^-$ & $6.0 \cdot 10^{-3}$ & $9.0 \cdot 10^{-3}$ \\
        $f_0^-$ & $3.9 \cdot 10^6$ & $4.2 \cdot 10^6$ \\
        $f_r^-$ & $1.0 \cdot 10^4$ & $1.0 \cdot 10^5$ \\
        $c_1$ & 0.01 & 0.9 \\
        $c_2$ & 0.01 & 0.6 \\
        $c_3$ & 0.3 & 2.2 \\
        \hline
    \end{tabular}
    \caption{Used bounds $\left\{ \boldsymbol{\Theta} \right\}$ for the optimized parameters.}
    \label{tab:my_bounds}
\end{table}

\noindent In order the optimizer to work with variables with the same order or magnitude (otherwise we manage variables with up to 9 order of magnitude difference) we apply an affine transformation which maps to original parameter values to a natural coordinate system $\xi_i$ that ranges from 0 to 1. This is done by:

\begin{equation}
    \xi_i = \frac{\boldsymbol{\Theta}_i - \left\{ \boldsymbol{\Theta} \right\}_{i, 0}}{\left\{ \boldsymbol{\Theta} \right\}_{i, 1} - \left\{ \boldsymbol{\Theta} \right\}_{i, 0}}
\end{equation}

Next, we use the cost function defined in Eq. \eqref{cost_L} to be optimized. Since the VL of 26 strain histories was previously run, the internal work ($W_{int,k}$) for each $n_{VL}$ are known, the optimizer will update the natural parameters values $\boldsymbol{\xi}$ to minimize the value of $\mathcal{L}(\boldsymbol{\xi}(\boldsymbol{\Theta}))$. Each time that the parameters are updated, the parametrized constitutive law reproduces the strain-stress history and then it is transformed to a strain-internal work one; then, the cost function is evaluated.

To avoid physically unacceptable combinations of parameters, we must inform the trust-region method with a set of linear and non-linear constraints. In this work, 4 constraints were added:
\begin{enumerate}
    \item $f_p^- > f_0^-$ (linear).
    \item $\varepsilon_p^- > f_0^- / E$ (linear).
    \item $l_{ch, RVE} - 2\,E\,G_f^+ / (f_t^+)^2 < 0$ (avoid tensile constitutive snap-back, nonlinear).
    \item $\frac{G_f^- / l_{ch,RVE} - G_{c1}}{\mathcal{G}_{Bezier} - G_{c1}} > 0$ (avoid compressive constitutive snap-back, nonlinear, see \cite{Petracca2017}).
\end{enumerate}

Next, we can finally execute the optimizer \texttt{scipy.optimize.minimize(-)} as it is described in Alg. \ref{algorithm:machine_learning_optimization_procedure} until convergence. After the required iterations, the algorithm converged to the following material properties, after converting to the real units:

\begin{equation}
\setcounter{MaxMatrixCols}{20}
    \begin{bmatrix}
        E \\ \nu \\ f_0^+ \\ G_f^+ \\ k_b \\ \kappa \\ G_f^- \\ f_p^- \\ \varepsilon_p^- \\ f_0^- \\ f_r^- \\ c_1 \\ c_2 \\ c_3
    \end{bmatrix}
    =     \begin{bmatrix}
        4.46701076 \cdot 10^9 \text{ (Identified in Section 3.4.1)}\\
        0.21639363 \text{ (Identified in Section 3.4.1)}\\
        2.6 \cdot 10^5 \\
        1.0 \cdot 10^3 \\
        1.15 \\
        2.7650 \cdot 10^{-7} \\
        8.0332 \cdot 10^2 \\
        1.0 \cdot 10^7 \\
        6.1007 \cdot 10^{-3} \\
        3.9874 \cdot 10^6 \\
        1.0 \cdot 10^4 \\
        4.9547 \cdot 10^{-1} \\
        0.6 \\
        2.2
    \end{bmatrix}
\end{equation}

\begin{figure}[htbp]
    \centering
    \begin{subfigure}[b]{0.3\textwidth}
        \centering
        \includegraphics[width=\textwidth]{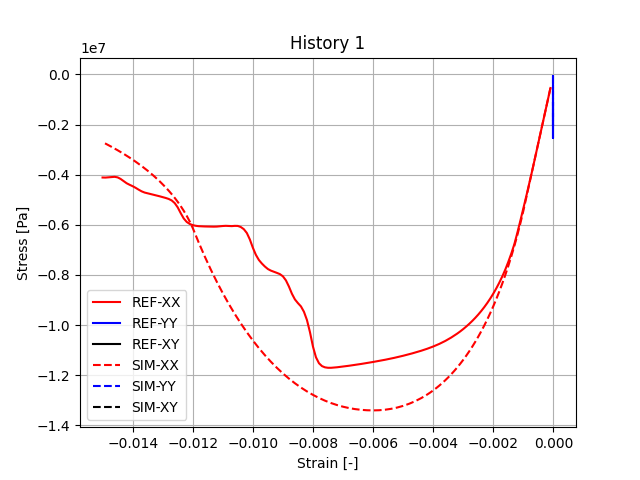}
        \caption{Case 1}
        \label{fig:subfig1}
    \end{subfigure}
    \hfill
    \begin{subfigure}[b]{0.3\textwidth}
        \centering
        \includegraphics[width=\textwidth]{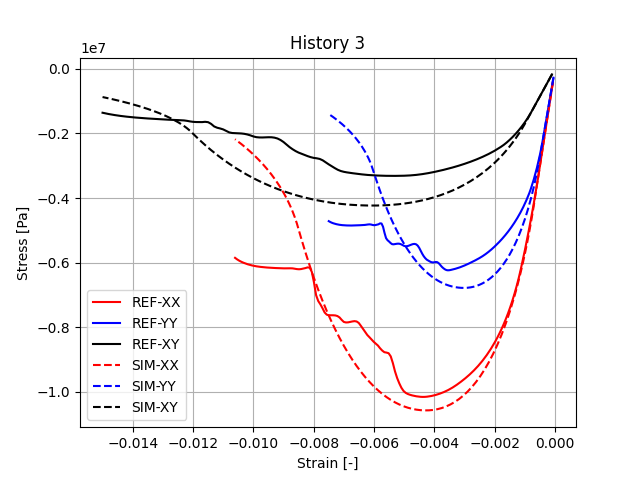}
        \caption{Case 3}
        \label{fig:subfig2}
    \end{subfigure}
    \hfill
    \begin{subfigure}[b]{0.3\textwidth}
        \centering
        \includegraphics[width=\textwidth]{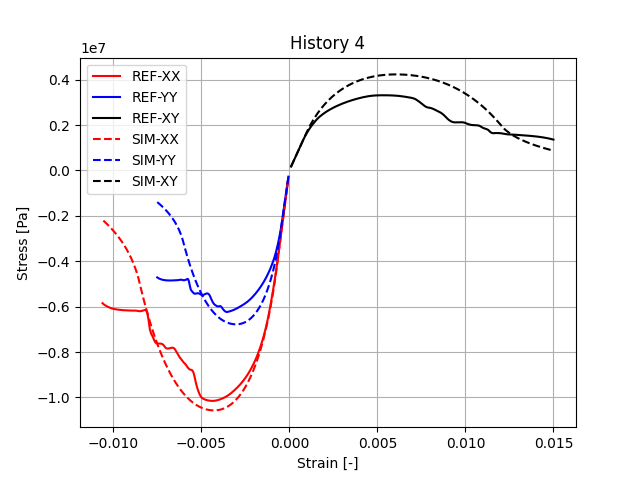}
        \caption{Case 4}
        \label{fig:subfig3}
    \end{subfigure}
    \vfill
    \begin{subfigure}[b]{0.3\textwidth}
        \centering
        \includegraphics[width=\textwidth]{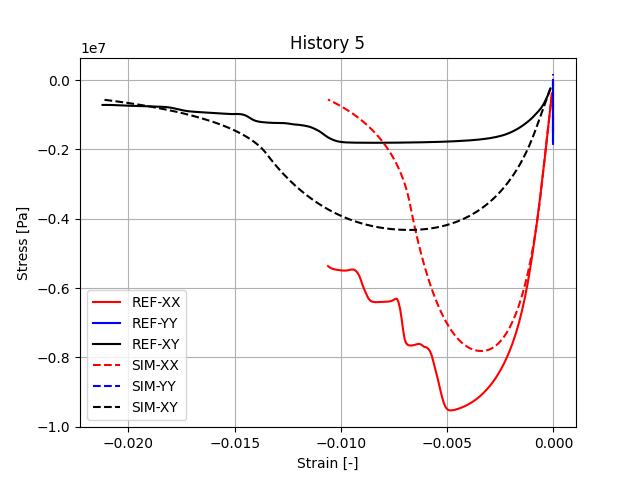}
        \caption{Case 5}
        \label{fig:subfig4}
    \end{subfigure}
    \hfill
    \begin{subfigure}[b]{0.3\textwidth}
        \centering
        \includegraphics[width=\textwidth]{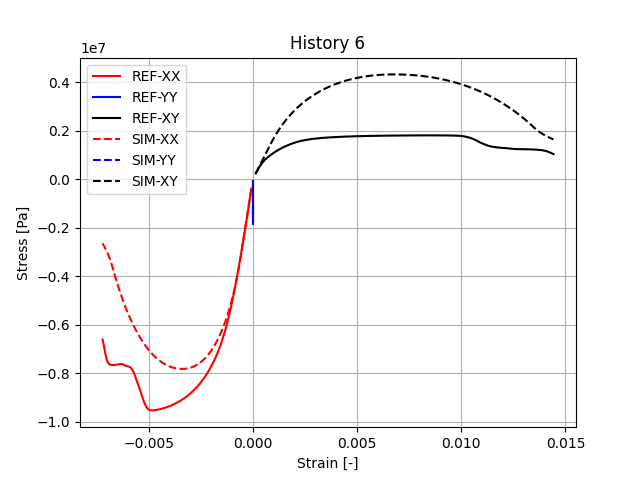}
        \caption{Case 6}
        \label{fig:subfig5}
    \end{subfigure}
    \hfill
    \begin{subfigure}[b]{0.3\textwidth}
        \centering
        \includegraphics[width=\textwidth]{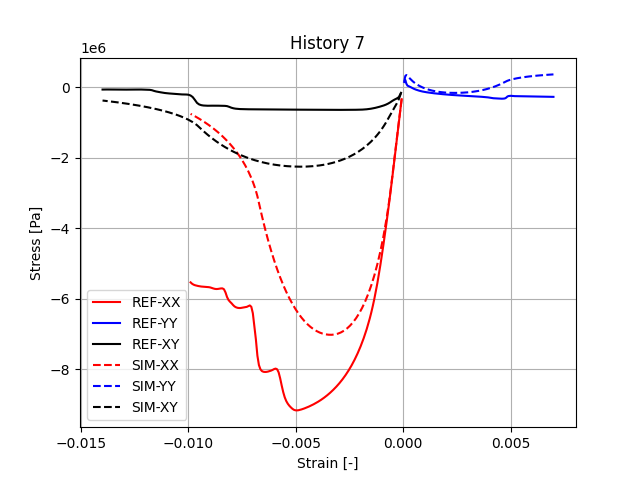}
        \caption{Case 7}
        \label{fig:subfig6}
    \end{subfigure}
    \vfill
    \begin{subfigure}[b]{0.3\textwidth}
        \centering
        \includegraphics[width=\textwidth]{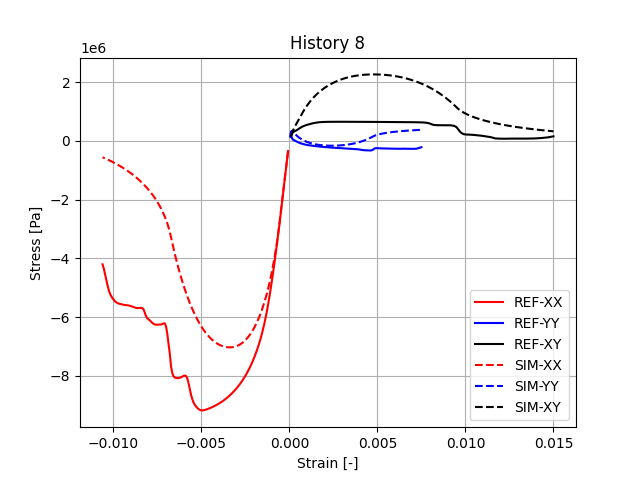}
        \caption{Case 8}
        \label{fig:subfig7}
    \end{subfigure}
    \hfill
    \begin{subfigure}[b]{0.3\textwidth}
        \centering
        \includegraphics[width=\textwidth]{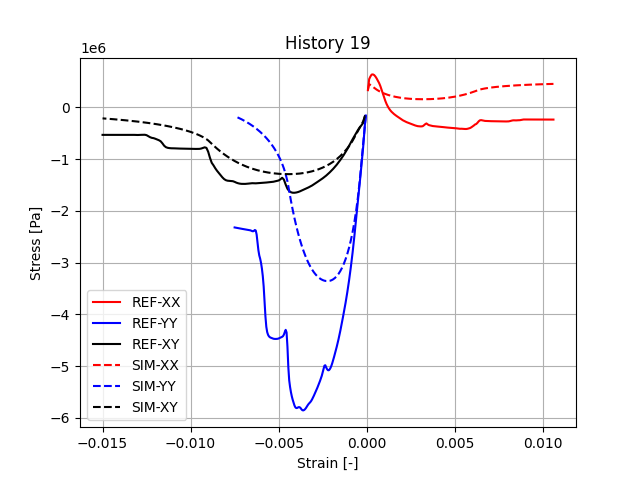}
        \caption{Case 19}
        \label{fig:subfig8}
    \end{subfigure}
    \hfill
    \begin{subfigure}[b]{0.3\textwidth}
        \centering
        \includegraphics[width=\textwidth]{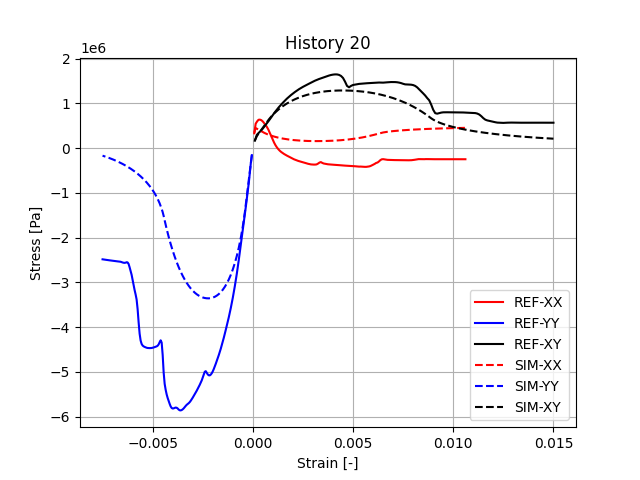}
        \caption{Case 20}
        \label{fig:subfig9}
    \end{subfigure}
 \centering
 \vfill
\begin{subfigure}[b]{0.3\textwidth}
        \centering
        \includegraphics[width=\textwidth]{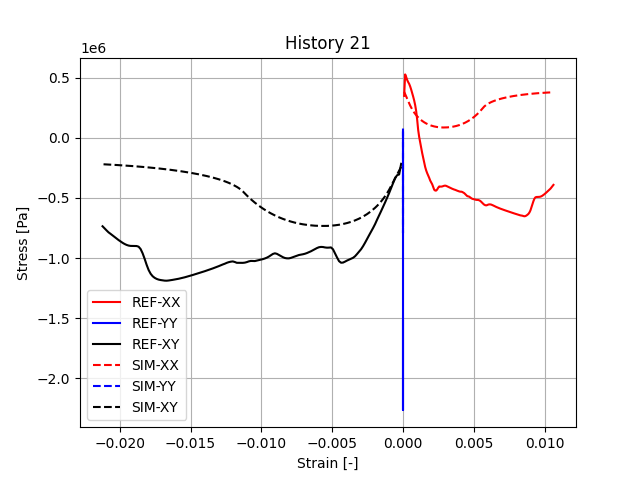}
        \caption{Case 21}
        \label{fig:subfig10}
    \end{subfigure}
    \hfill
    \begin{subfigure}[b]{0.3\textwidth}
        \centering
        \includegraphics[width=\textwidth]{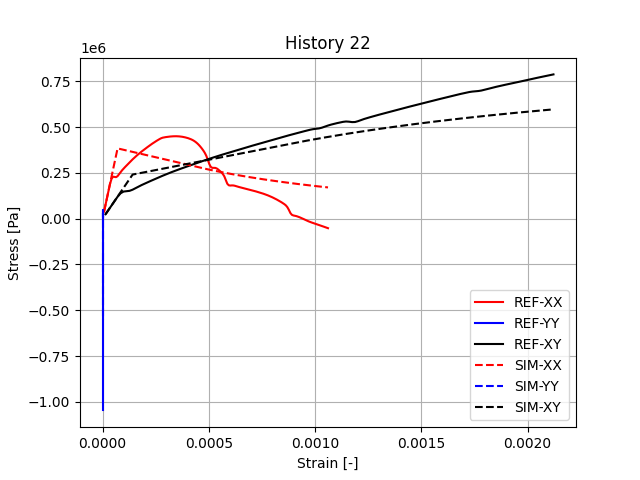}
        \caption{Case 22}
        \label{fig:subfig10}
    \end{subfigure}
    \hfill
\begin{subfigure}[b]{0.3\textwidth}
        \centering
        \includegraphics[width=\textwidth]{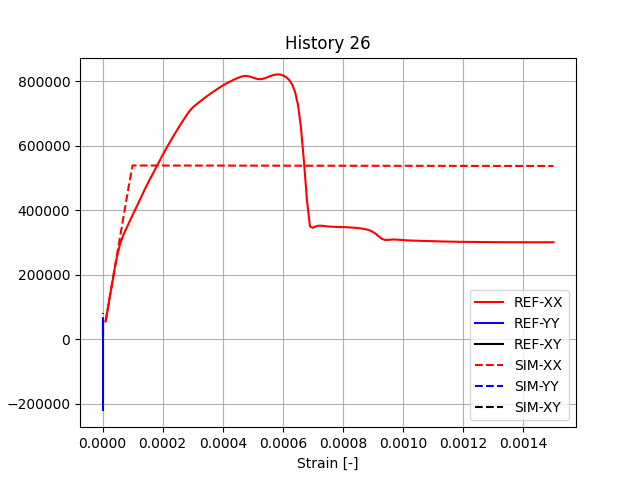}
        \caption{Case 26}
        \label{fig:subfig10}
    \end{subfigure}
    \caption{Selection of 10 strain history cases, comparison of the stress components from the reference VL results (solid lines) and the homogenized model ones (dashed lines). XX, YY and XY stands for each of the stress vector components.}
    \label{fig:results_VL_stress}
\end{figure}

Fig. \ref{fig:results_VL_stress} shows a detailed comparison of the strain-stress histories for 10 of the 26 cases run in the VL obtained from the high fidelity micro model (RVE) and the ones coming from the homogenized and optimized macro model. Each color represents a different stress vector component ($\sigma_x, \sigma_y, \sigma_{xy}$), being the solid lines the micro and dashed the macro model results. It can be seen how the optimized parameters are able to reasonably match the macro model results to the reference ones. Cases 1, 3, 4 are clearly well represented by the macro model. However, in other cases in which the micro mechanics on the high fidelity model is more relevant, the macro model finds it difficult to accurately reproduce it, this can be studied in cases 7, 19 and 20, for example. Fortunately, the optimization algorithm aims at minimizing the overall difference in terms of internal work and not individual strain-stress histories; hence, the resulting set of optimized material properties, as one can see in the next section, sufficiently reproduces the macro behaviour of this kind of masonry structures.

\begin{figure}[htbp]
    \centering
    \begin{subfigure}[b]{0.3\textwidth}
        \centering
        \includegraphics[width=\textwidth]{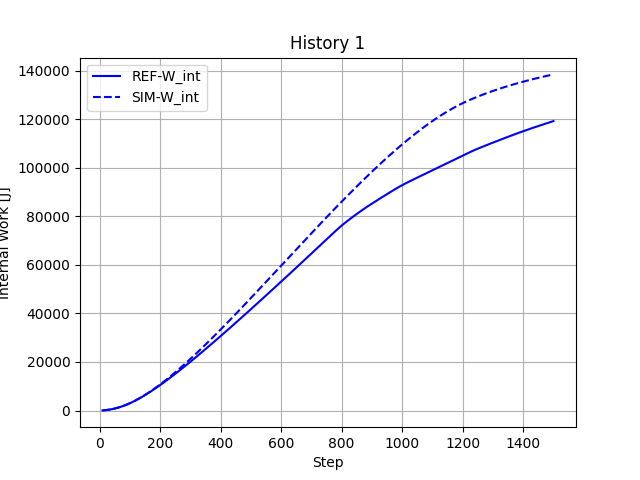}
        \caption{Case 1}
    \end{subfigure}
    \hfill
    \begin{subfigure}[b]{0.3\textwidth}
        \centering
        \includegraphics[width=\textwidth]{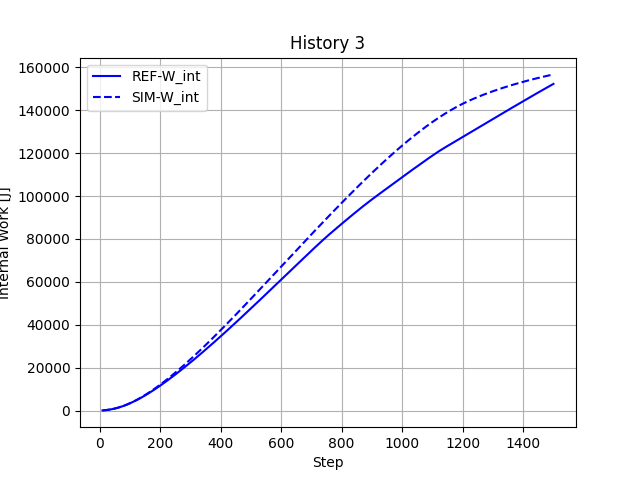}
        \caption{Case 3}
        \label{fig:subfig12}
    \end{subfigure}
    \hfill
    \begin{subfigure}[b]{0.3\textwidth}
        \centering
        \includegraphics[width=\textwidth]{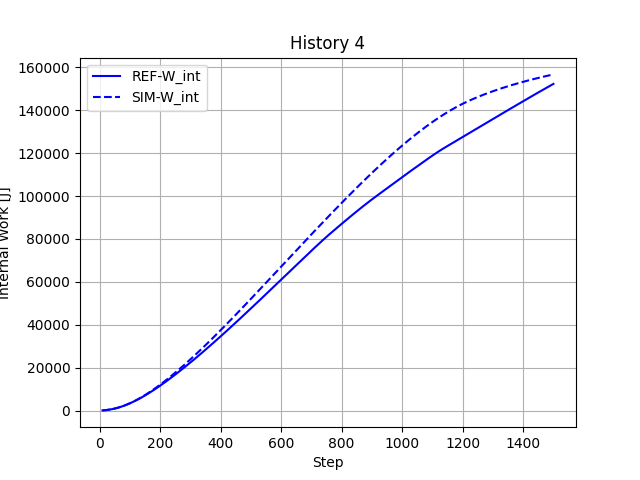}
        \caption{Case 4}
        \label{fig:subfig13}
    \end{subfigure}
    \vfill
    \begin{subfigure}[b]{0.3\textwidth}
        \centering
        \includegraphics[width=\textwidth]{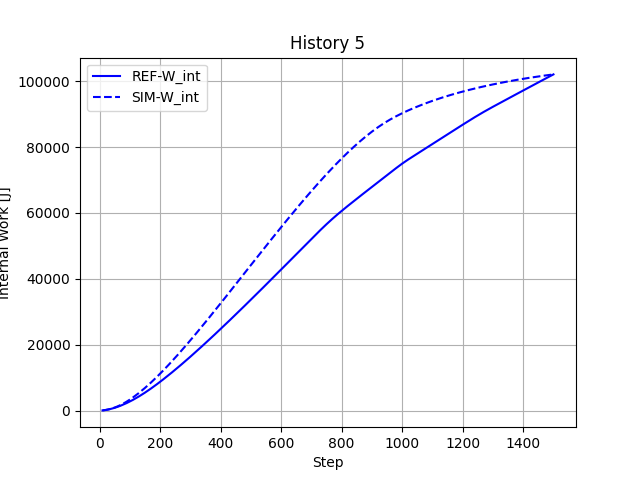}
        \caption{Case 5}
        \label{fig:subfig14}
    \end{subfigure}
    \hfill
    \begin{subfigure}[b]{0.3\textwidth}
        \centering
        \includegraphics[width=\textwidth]{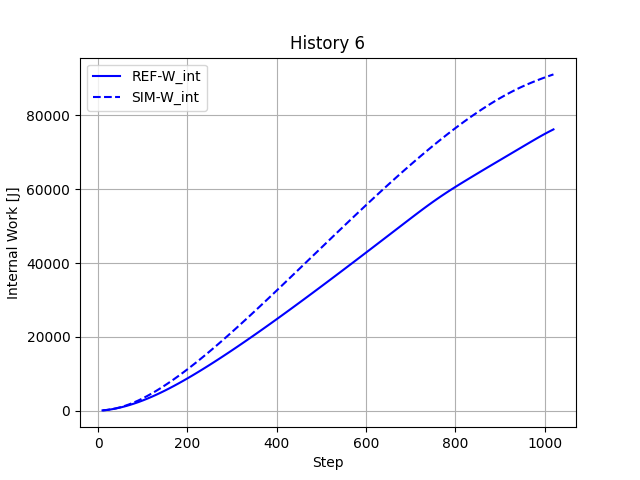}
        \caption{Case 6}
        \label{fig:subfig15}
    \end{subfigure}
    \hfill
    \begin{subfigure}[b]{0.3\textwidth}
        \centering
        \includegraphics[width=\textwidth]{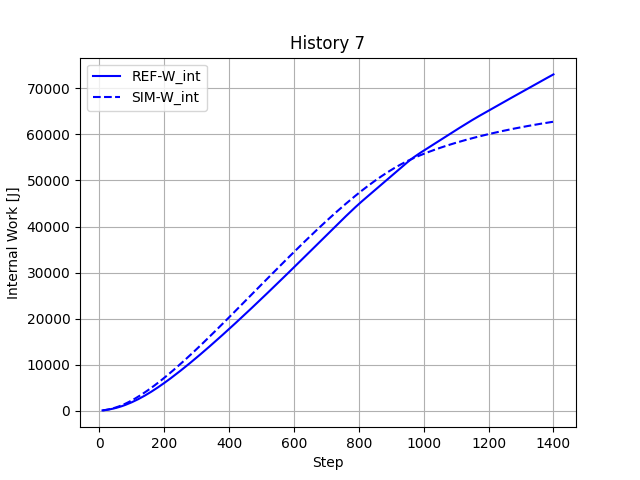}
        \caption{Case 7}
        \label{fig:subfig16}
    \end{subfigure}
    \vfill
    \begin{subfigure}[b]{0.3\textwidth}
        \centering
        \includegraphics[width=\textwidth]{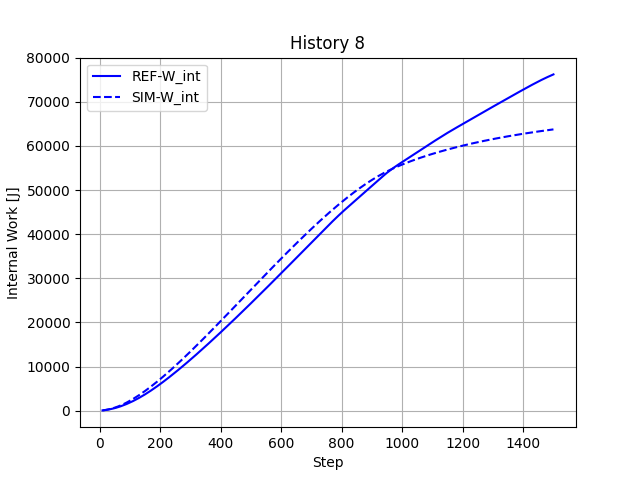}
        \caption{Case 8}
        \label{fig:subfig17}
    \end{subfigure}
    \hfill
    \begin{subfigure}[b]{0.3\textwidth}
        \centering
        \includegraphics[width=\textwidth]{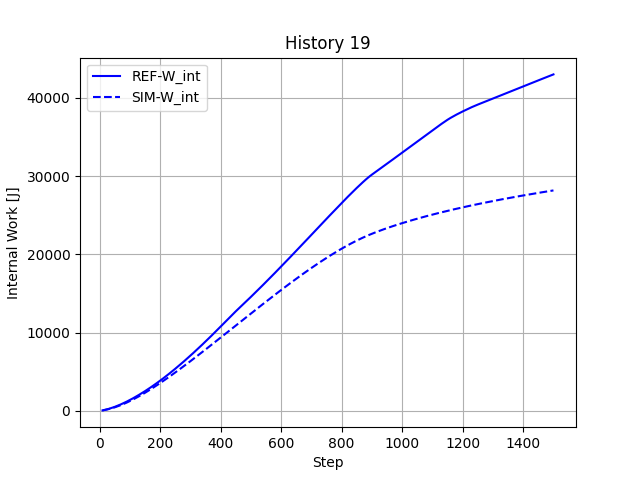}
        \caption{Case 19}
        \label{fig:subfig18}
    \end{subfigure}
    \hfill
    \begin{subfigure}[b]{0.3\textwidth}
        \centering
        \includegraphics[width=\textwidth]{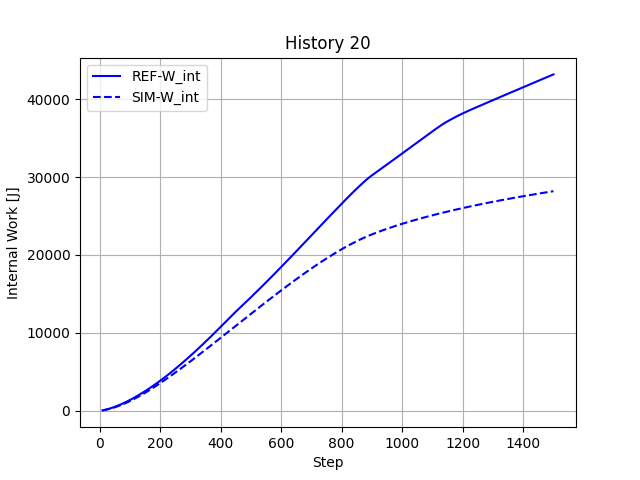}
        \caption{Case 20}
        \label{fig:subfig19}
    \end{subfigure}
 \centering
 \vfill
\begin{subfigure}[b]{0.3\textwidth}
        \centering
        \includegraphics[width=\textwidth]{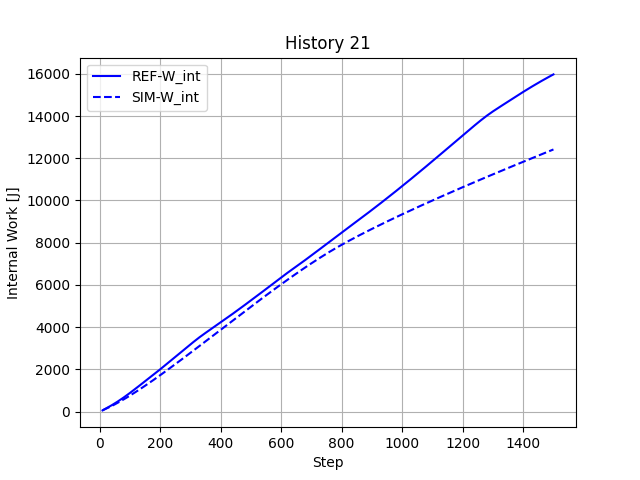}
        \caption{Case 21}
        \label{fig:subfig1111110}
    \end{subfigure}
    \hfill
    \begin{subfigure}[b]{0.3\textwidth}
        \centering
        \includegraphics[width=\textwidth]{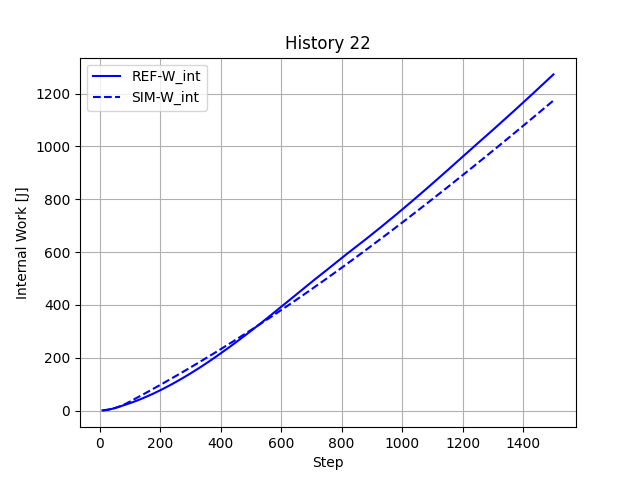}
        \caption{Case 22}
    \end{subfigure}
    \hfill
\begin{subfigure}[b]{0.3\textwidth}
        \centering
        \includegraphics[width=\textwidth]{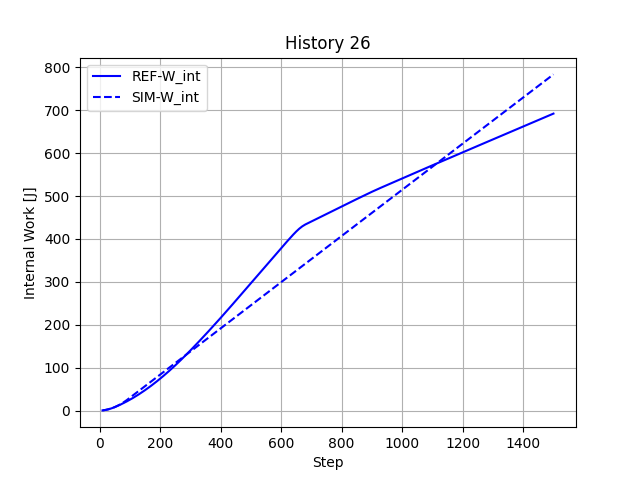}
        \caption{Case 26}
    \end{subfigure}
    \caption{Selection of 10 history cases, comparison of the internal work from the reference VL results at each time increment (solid lines) and the homogenized model ones (dashed lines).}
    \label{fig:results_VL_Work}
\end{figure}

Additionally, in Fig. \ref{fig:results_VL_Work} the time evolution of the dissipated internal work is depicted, whose last value of each load case is the main control value for optimization. Since the internal work is a postprocess from the strain-stress curves, similar conclusions can be drawn. Initially, the elastic internal work are identical in both cases since the elastic orthotropy is perfectly captured. Then, as soon as damage starts, the degradation mechanisms, which depend on the optimizable parameters, may differ. In this case, the optimizer is able to find a compromise solution in which, overall, the difference in terms of internal work is balanced and as close as the macro model limitations allows.

\subsection{Post machine learning application} \label{sec:application_post_ml}

\subsubsection{Numerical analyses}

\begin{figure}[h]
\centering
	\includegraphics[width=0.8\linewidth]{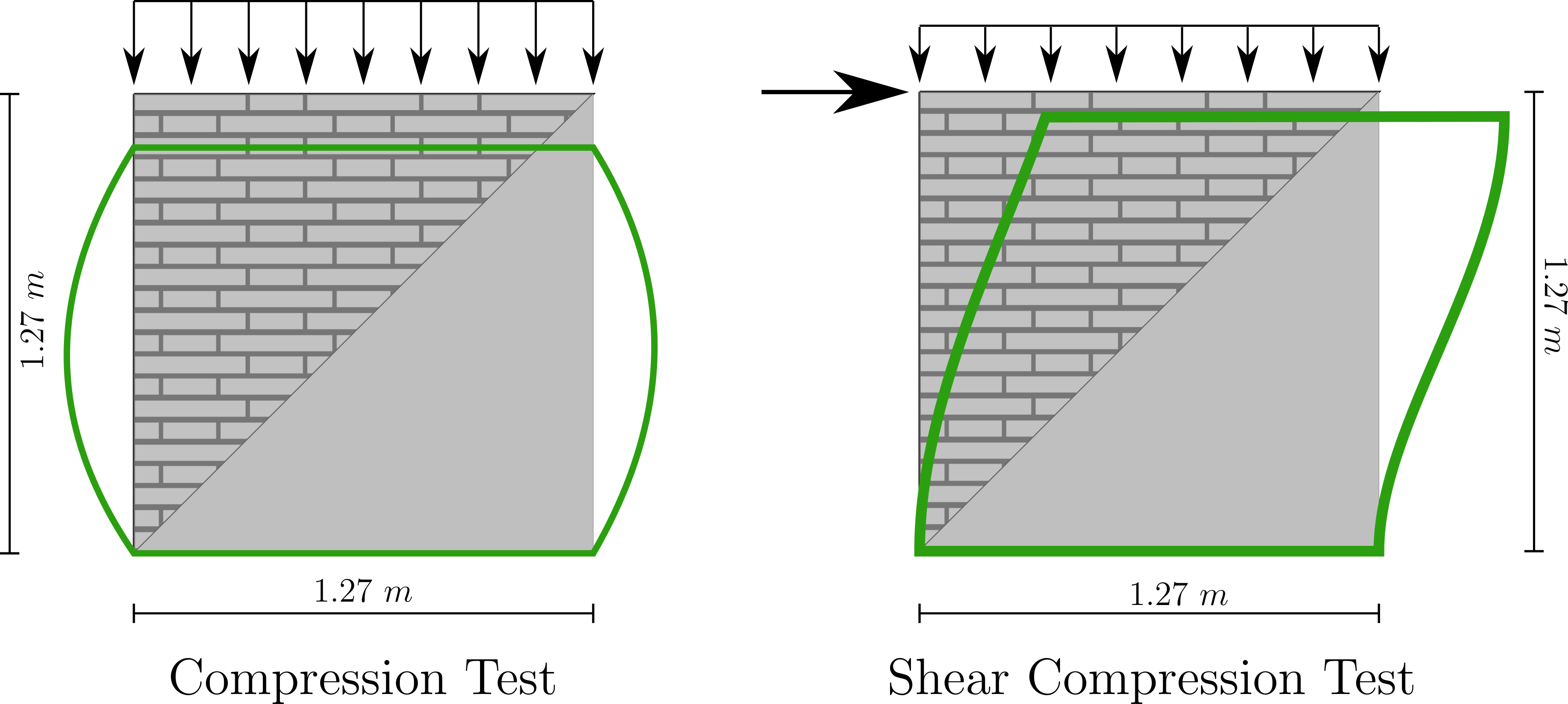} 
    \caption{Schematic views of the models of the compression and the shear compression tests. A numerical micro model, distinguishing between mortar joints and bricks units, and a smeared macro model are the substance of the analyses \cite{PhilippThesis}.}
\label{fig:machine_learning_application_post_ml_numerical_test_overview}
\end{figure}

Subject of the numerical analyses are three 2D micro and a macro models of the same masonry wall. The micro model wall respects the same geometrical distribution of  bricks (Flemish bond) as utilized for the RVE of the virtual laboratory. However, the dimensions of the wall are $1.27m \times 1.27m$, which is larger than the dimensions of the RVE. Each of the two walls is opposed to two different boundary conditions. The first analysis is a compression test and the second and third, shear compression tests. The walls are loaded by incrementally applying displacements on top of the wall until failure. Fig. \ref{fig:machine_learning_application_post_ml_numerical_test_overview} shows the concepts for both the numerical tests. It also indicates that either a micro modeling or a smeared macro model approach is applied to this analysis.

\paragraph{Compression test}
The finite element nodes of the base of the wall are fixed for displacements in any direction. The load is applied by monotonically increasing a vertical displacement to the top nodes of the wall's finite element model in order to perform the compression test. 

\paragraph{Shear compression tests} 
The finite element model for the shear compression tests is subjected to two loading stages. The first is an incrementally increasing vertical displacement analogously to the model of the compression test.  This vertical displacement moves up to a predefined value of $d_y = 0.09$ mm where no failure of the wall is caused. The second shear test increases the precompression by a 30\%. This vertical displacement is kept constant during the second step of the analysis. After having reached the vertical displacement, a monotonic increasing  horizontal top displacement is applied in order to burden the wall in shear. The horizontal displacement is then applied until reaching failure of the model.

A total number of six numerical analyses are performed in this work. Three of them are the micro model analyses of the compression and the shear compression tests, respectively. The material properties of the micro model components have been introduced in Table \ref{table:machine_learning_law_parameters_diagonal_compression} and are also applied here. The remaining three analyses are the macro model analysis of both the compression and the shear compression tests. The smeared material properties of the macro models have been predicted in the previous sections in terms of optimal fitting of the parameters $\bm{\Theta}$ of the constitutive model $\Psi$.

All the analysis are calculated by utilizing the open source finite element software \textsc{kratos-multiphysics}. The analyses at macro scale utilize the novel implemented constitutive law introduced in Section \ref{sec:post_ml_cl}. It applies the transformation matrix $\textbf{T}$ in order to map the strains and stresses from the anisotropic to the isotropic scale (and vice versa). 

\subsubsection{Analyses results and comparison}

\paragraph{Compression test}

\begin{figure}[h]
\centering
	\includegraphics[width=0.5\linewidth]{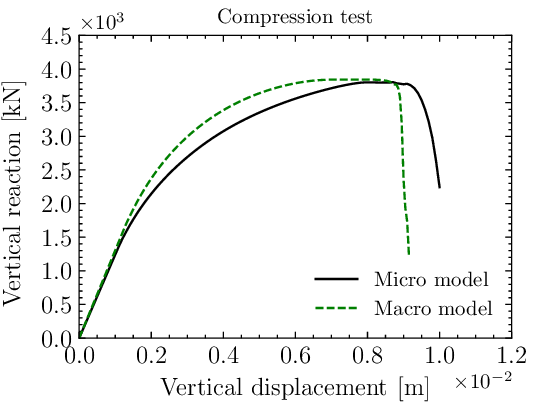} 
    \caption{Compression test: vertical reaction force vs. vertical displacement curves of the micro model and the macro model $\Psi$ analysis results.}
\label{fig:machine_learning_application_post_ml_compression_curves}
\end{figure}

The vertical reaction force of the three analyses was measured during the boundary application in order to compare the results. Fig. \ref{fig:machine_learning_application_post_ml_compression_curves} shows this force plotted against the applied vertical top displacement for the micro and the macro model analyses. All three tests remain in the linear range up to a top displacement of $d_y \approx 0.1$ cm. The initial stiffness of all tests is similar. After having reached the damage onset, material hardening starts for all analyses. The hardening progress can also be seen as approximately equal up to a top displacement of $d_y \approx 0.9$ cm. It can be seen that the maximum vertical reaction forces and the corresponding top displacements of the two models are indeed in good agreement.  {It is important to mention that in the macro model simulations, an enhanced accuracy mixed $u/\varepsilon$ FE \cite{Mixed1, Mixed2} has been used to avoid spurious mesh dependent results. This mixed FE solves an enriched set of degrees of freedom, \textit{i.e.} nodal displacements and strains are solved and compatibilized. This increases the number of unknowns (3 more in 2D and 5 in 3D) but the results are mesh independent since full compatibility of the strain field is achieved.}

\begin{figure}[htbp]
    \centering
    \begin{subfigure}[b]{0.45\textwidth}
        \centering
        \includegraphics[width=\textwidth]{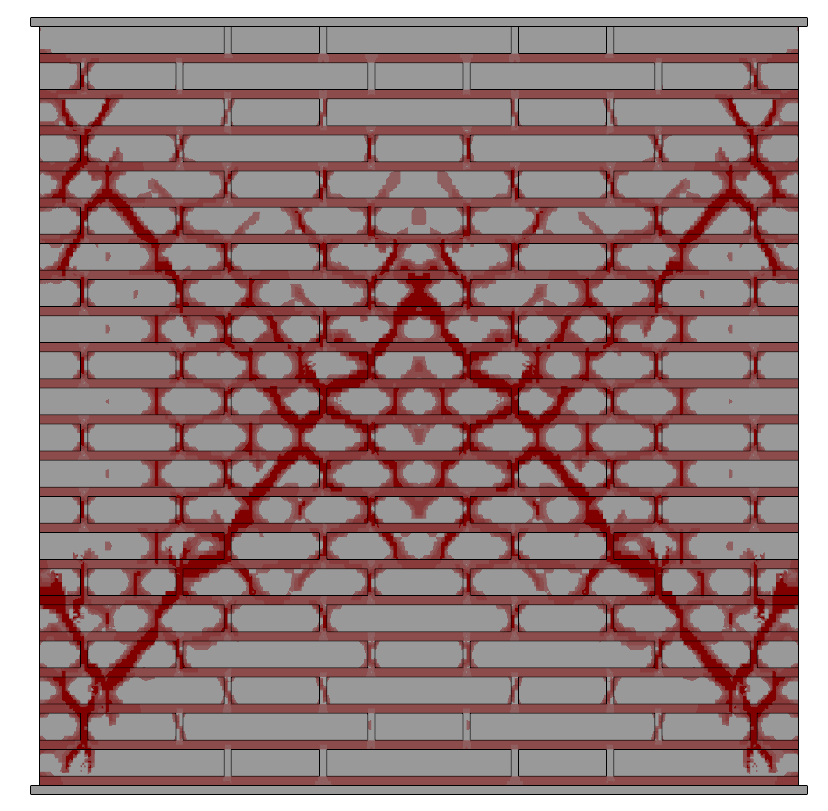}
        \caption{Micro model: damage field}
        \label{fig:subfig1}
    \end{subfigure}
    \hfill
    \begin{subfigure}[b]{0.45\textwidth}
        \centering
        \includegraphics[width=\textwidth]{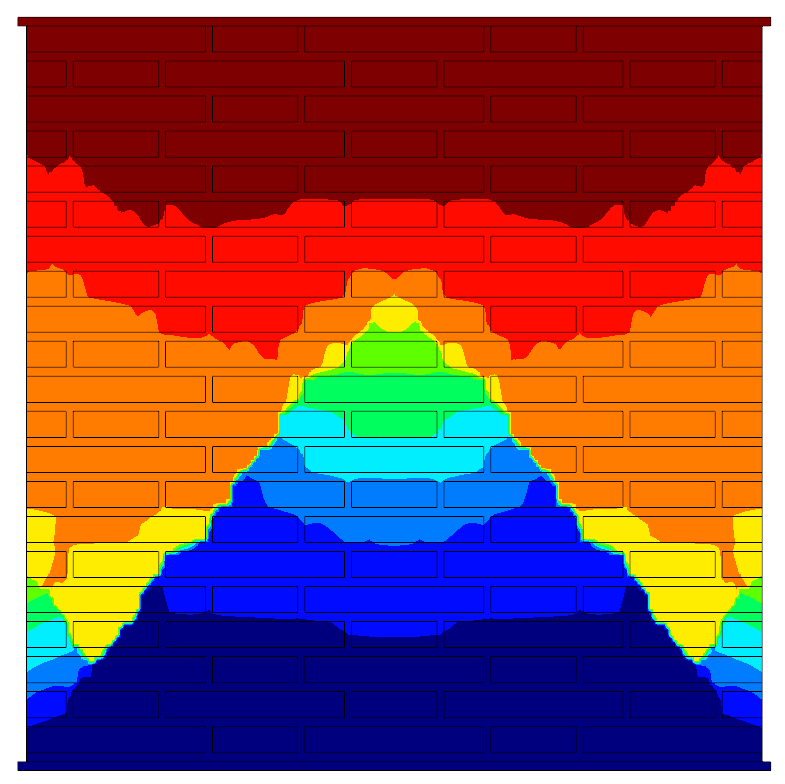}
        \caption{Micro model, displacements}
        \label{fig:subfig2}
    \end{subfigure}
    \vfill
    \begin{subfigure}[b]{0.45\textwidth}
        \centering
        \includegraphics[width=\textwidth]{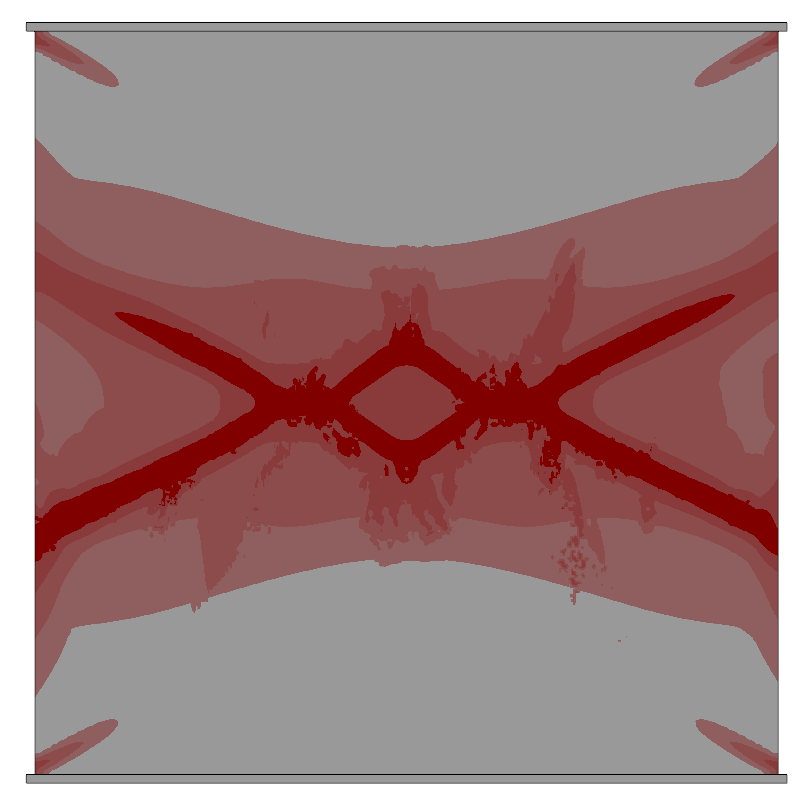}
        \caption{Macro model: damage field}
        \label{fig:subfig3}
    \end{subfigure}
    \hfill
    \begin{subfigure}[b]{0.45\textwidth}
        \centering
        \includegraphics[width=\textwidth]{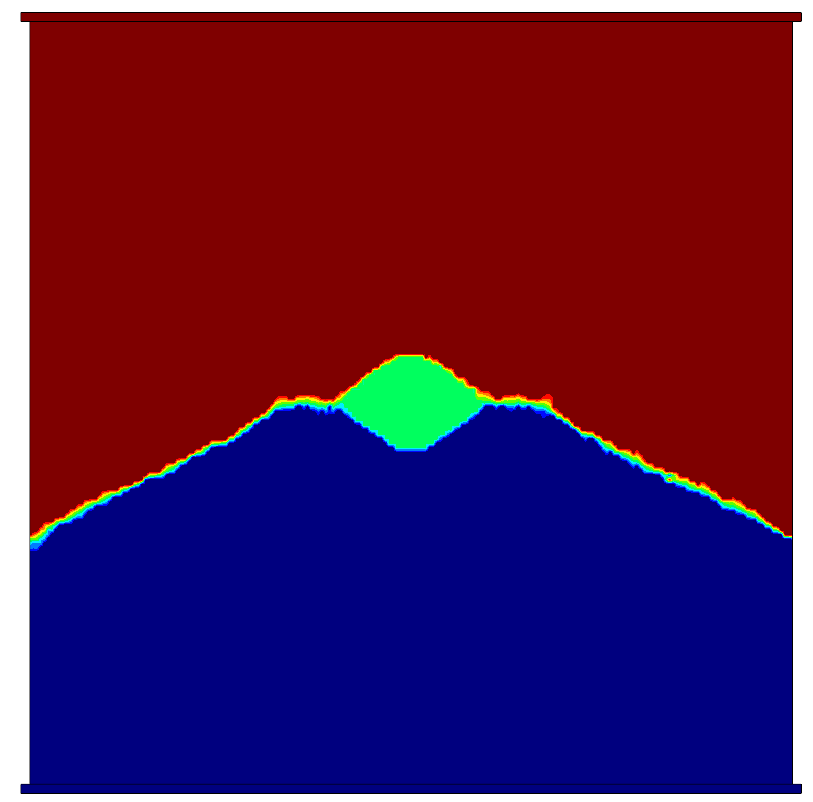}
        \caption{Macro model: displacements}
        \label{fig:subfig4}
    \end{subfigure}
    \caption{Compression test: damage patterns of the micro model and the macro model $\Psi$ analysis results, respectively. Demonstrating the contour plots of the total displacements ranging from 0 (blue) to 0.01 m (red).}
\label{fig:machine_learning_application_post_ml_compression_damage_plots}
\end{figure}

Fig. \ref{fig:machine_learning_application_post_ml_compression_damage_plots} shows the crack patterns of the compression tests in terms of the contour plots of the damage variable and total displacements. All plots are taken at a top displacement of $d_y=10.0$ mm. The crack pattern of the micro model analysis exhibits the hourglass shape which is typical for compression tests on brittle materials such as masonry. Both the macro model analyses are able to produce a crack pattern very similar to the one of the micro model. The results of the compression test have shown that both the macro models are appropriate models when compared with the compression test analysis of a micro model.

\paragraph{Shear compression tests}

\begin{figure}[h]
\centering
    \begin{subfigure}[b]{0.45\textwidth}
        \centering
        \includegraphics[width=\textwidth]{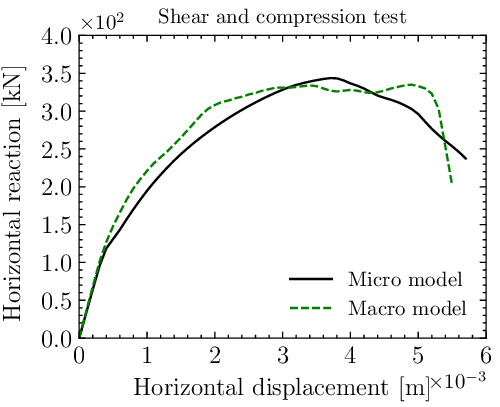}
        \caption{Shear compression}
        \label{fig:shear}
    \end{subfigure}
    \begin{subfigure}[b]{0.45\textwidth}
        \centering
        \includegraphics[width=\textwidth]{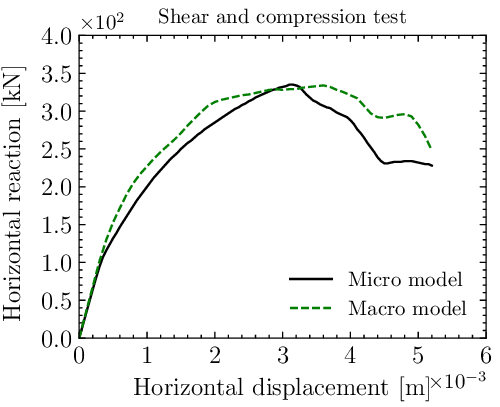}
        \caption{Shear $+30\%$ compression }
        \label{fig:shear_extra}
    \end{subfigure}
    \caption{Shear compression test: curves showing the horizontal reaction force vs. horizontal displacement of the micro model and the macro model $\Psi$ analysis results, respectively.}
\label{fig:machine_learning_application_post_ml_shearcompression_curves}
\end{figure}

Two shear compression tests were conducted, one adding an extra 30\% of compressive load. The comparison of the shear compression test applied to the micro and the macro models is based on the measurements of the horizontal reaction force $F_x$ and its corresponding horizontal displacement. Fig. \ref{fig:machine_learning_application_post_ml_shearcompression_curves} depicts the curves of the force plotted against the horizontal top displacement $d_x$ for the two analyses. The initial stiffness demonstrates an identical linear elastic behavior for both micro and macro models. A damage onset starts at a horizontal top displacement of $d_x \approx 0.5$ mm. Further increasing of the deformation leads to similar behaviors of the three models up to a reaction force of $F_x \approx 350$ kN. From there on, both the micro and macro models exhibit a clear softening behaviour.

\begin{figure}[htbp]
    \centering
    \begin{subfigure}[b]{0.45\textwidth}
        \centering
        \includegraphics[width=\textwidth]{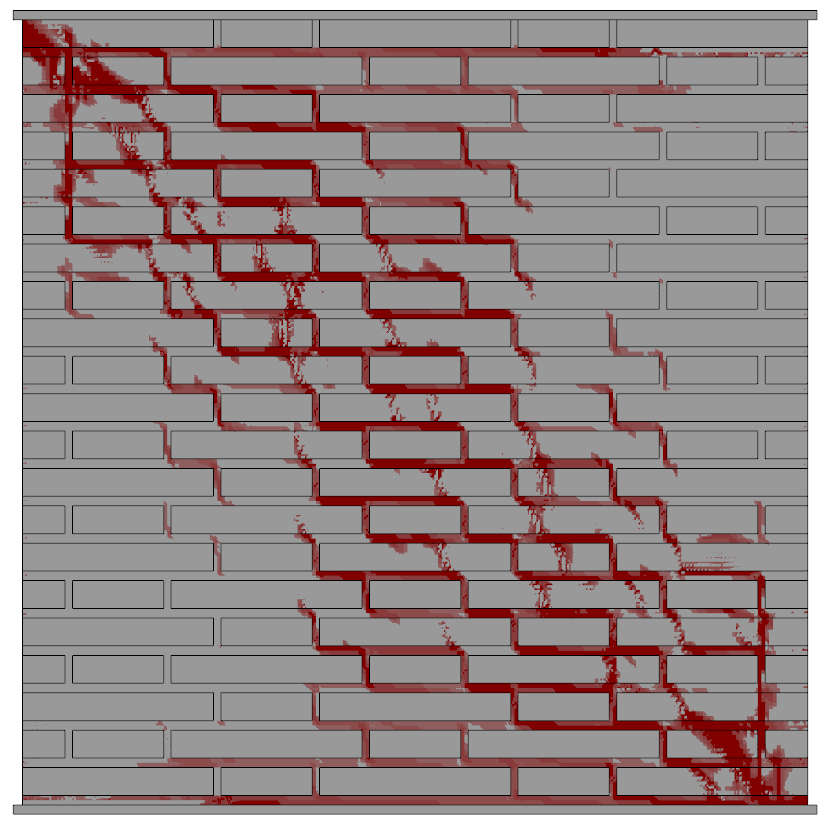}
        \caption{Micro model: damage field}
    \end{subfigure}
    \hfill
    \begin{subfigure}[b]{0.45\textwidth}
        \centering
        \includegraphics[width=\textwidth]{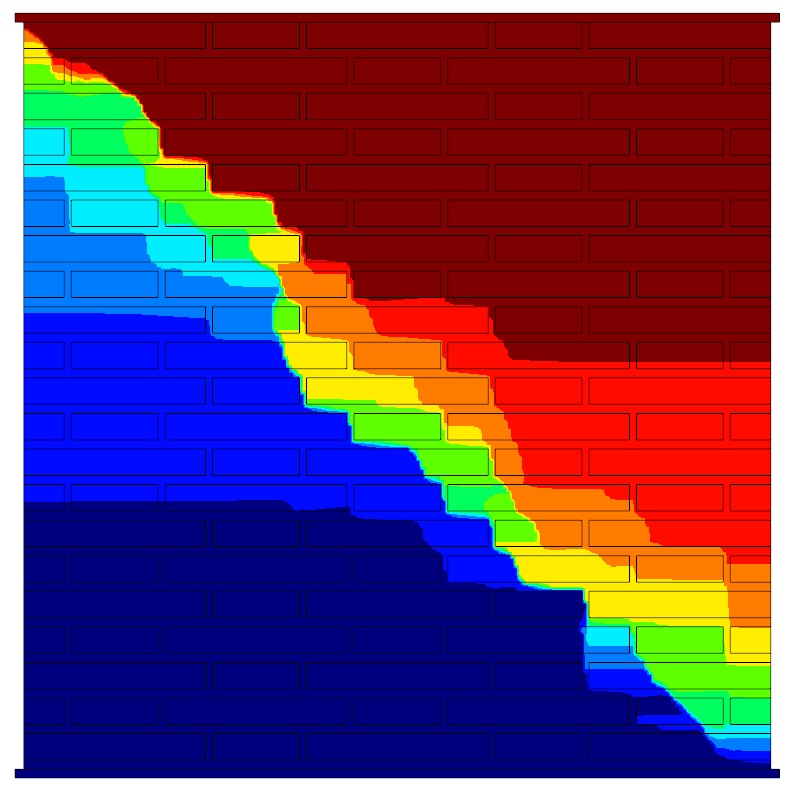}
        \caption{Micro model, displacements}
    \end{subfigure}
    \vfill
    \begin{subfigure}[b]{0.45\textwidth}
        \centering
        \includegraphics[width=\textwidth]{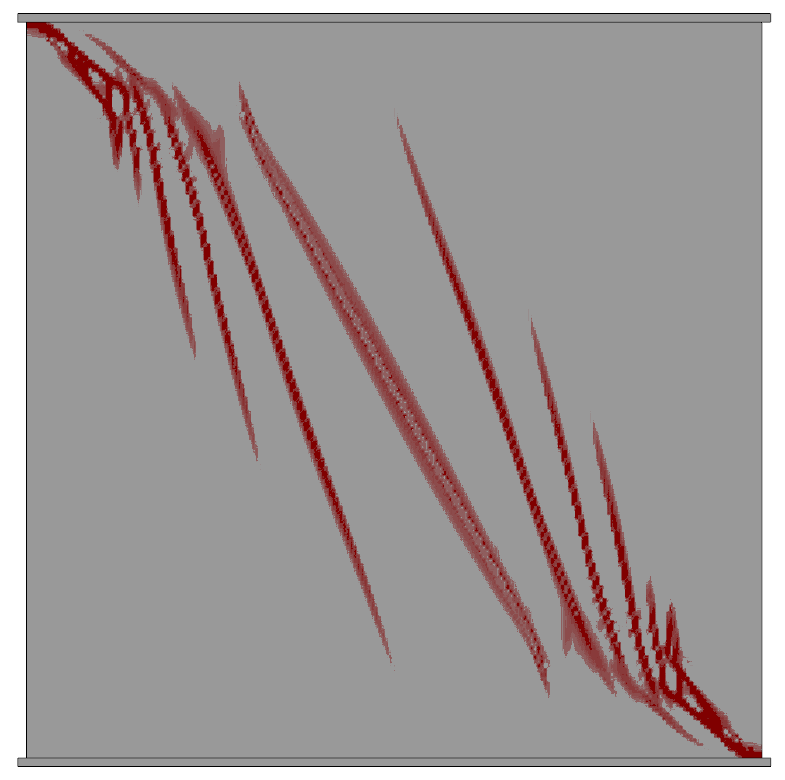}
        \caption{Macro model: damage field}
    \end{subfigure}
    \hfill
    \begin{subfigure}[b]{0.45\textwidth}
        \centering
        \includegraphics[width=\textwidth]{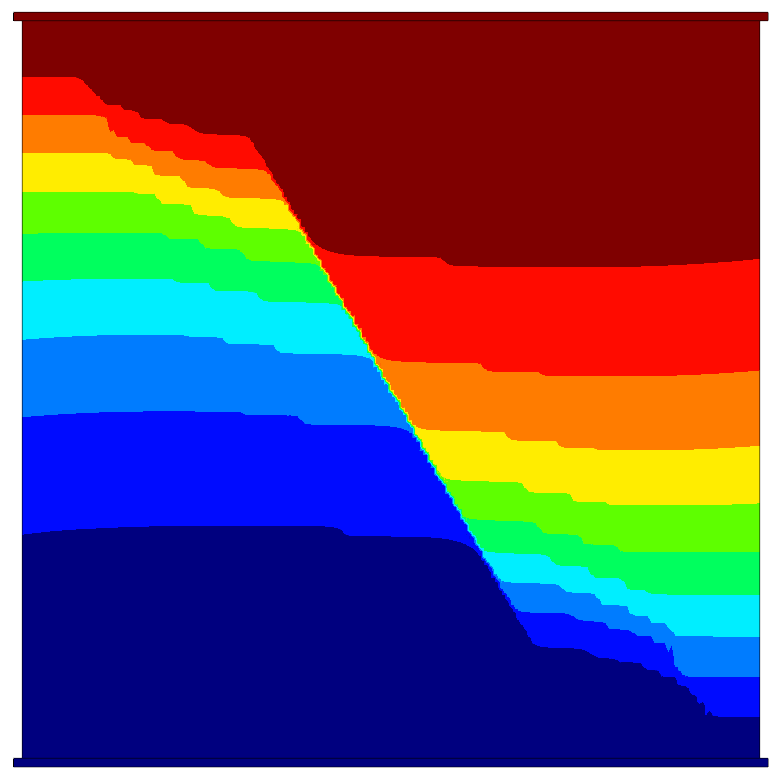}
        \caption{Macro model: displacements}
    \end{subfigure}
    \caption{Shear compression test: crack patterns of the micro model and the macro model $\Psi$, and the contour plots of the  displacement at a horizontal top displacement of $d_x = 6.0$ mm.}
\label{fig:machine_learning_application_post_ml_shearcompression_damage_plots}
\end{figure}

Fig. \ref{fig:machine_learning_application_post_ml_shearcompression_damage_plots} shows the contour plots of the damage field and the global displacement at a horizontal top displacement of $d_x = 6.0$ mm for the results of the micro and the macro model analyses, respectively. Since similar results are obtained in the second extra compression case, we avoid being redundant are not included in here. The damage contour plot of the micro model demonstrates a diagonal crack that passes through the mortar joints and the brick units. Such a pattern is typical for shear compression tests of masonry walls \cite{Petracca_thesis}. This single diagonal crack leads to ultimate failure of the micro model. The crack pattern of macro model $\Psi$, shows a series of diagonal cracks that are initiated in a sequential way until the central dominant one finally develops. This crack leads to failure of the model at a horizontal top displacement very similar to the one of the micro model. However, the inclination of the diagonal main crack is not 45º oriented in the macro model like it is shown in the micro simulation.

Overall, the numerical analysis results of the compression and the shear compression tests show that the macro model $\Psi$ is able to represent the micro model results very accurately. This is shown in both the comparison of the force displacement curves  and in the crack evolution on the other hand.

\section{Conclusions and future work} \label{sec:conclusions}

This research has presented the definition of a machine learning based technique able to homogenize a composite material like masonry. The smeared properties, within the framework of a macro modelling approach, are presented in a single damage constitutive model. In comparison to conventional homogenization techniques, the novel procedure disconnects micro and macro analysis scales while homogenizing the material, with important benefits in terms of computational efficiency.

Setting a virtual laboratory allows one to produce numerical results  that can be utilized for big data analysis tools, such as machine learning or artificial intelligence. This approach is clearly promoting the idea of assessing the mechanical properties of material by the use of directly processed data. In this regard, a structural simulation performed in the virtual laboratory produced data that accurately represent the heterogeneous material behavior of masonry and could so be used for training of a specific machine learning model.    {It is important to mention that there is no need for a large dataset for accurately estimating the macro properties of masonry. In this paper, 26 high fidelity RVE simulations were calculated with varying strain field aiming to cover any biaxial strain field that may occur. The novel utilization of strain-internal work histories as a measure of the cost function to be optimized has demonstrated to be a more manageable and meaningful than raw strain-stress relationships. After a prior isotropization of the data, the optimizer is able to efficiently estimate the best set of parameters for the macro model, even with a set of 12 parameters.}

The application and discussion of the developed homogenization tool to a Flemish bond masonry wall has shown results that are in very good agreement with those obtained from micro modeling, with a highly reduced computational cost.  {In this work, the application examples were conducted: one pure compression and two different shear compression tests. All  of them showed a very good agreement in quantitative force-displacement terms and fair similarity in crack development. }

Therefore, the presented machine learning based homogenization technique has shown to represent accurately the complex behavior of in-plane loaded masonry walls.

{Regarding the future work, in this paper we have optimized the parameters of a macro model to accurately predict the behaviour of a high fidelity virtual laboratory. In the following works, we aim to generalize this methodology to be able to optimize macro models whose reference information comes from real physical experiments or obtained from Digital Image Correlation (DIC) displacement/strain fields. In the first case the computation of internal work is not possible but, on the other hand, the optimizer can use the external one (tractions and boundary displacements), which they should be identical. In the case of DIC, the model will be upgraded by means of the Virtual Field Method (VFM) \cite{vfm}; this will enable the material and parameter identification of complex geometries, materials and loading situations, usually unfeasible to perform in standard experiments.}

\section{Acknowledgements}

The authors gratefully acknowledge the financial support from the Ministry of Science, Innovation and Universities of the Spanish Government (MCIU), the State Agency of Research (AEI) as well as that of the ERDF (European Regional Development Fund) through the project SEVERUS (Multilevel evaluation of seismic vulnerability and risk mitigation of masonry buildings in resilient historical urban centres, ref. Num. RTI2018-099589-B-I00). The authors gratefully acknowledges the AGAUR agency of the Generalitat de Catalunya for the financial support of one predoctoral grant. \\
Acknowledge the support received by
the Severo Ochoa Centre of Excellence (2019-2023) under the grant CEX2018-000797-S funded by 
 CIN/AEI/10.13039/501100011033. \\
Prof. Alejandro Cornejo is a Serra Húnter fellow.

\bibliographystyle{unsrt}  


\end{document}